\documentclass[lettersize,journal]{IEEEtran} 
\usepackage{cite}
\usepackage[colorlinks]{hyperref}
\usepackage{mathrsfs}
\usepackage{amsthm}
\usepackage{etoolbox} \makeatletter \patchcmd{\@makecaption} {\scshape} {} {} {} \makeatother
\usepackage{diagbox}
\usepackage[cmex10]{amsmath}
\usepackage{mathtools}
 \usepackage{siunitx}
\usepackage{array}
\usepackage{eqparbox}
\usepackage{bm}

\usepackage[table,dvipsnames]{xcolor}
\usepackage{multicol,booktabs,tabularx}

\usepackage{cases}
\usepackage{algorithm}
\usepackage{paralist}
\usepackage{algpseudocode}

\usepackage{graphicx}
\usepackage{subfigure}
\usepackage{epstopdf}
\usepackage{epsfig}
\usepackage{amssymb}
\usepackage{array}
\usepackage{multirow}

\begin{document}

\title{SSNet: Flexible and robust channel extrapolation for fluid antenna systems enabled by an self-supervised learning framework}

\author{Yuan Gao, Yiming Liu, Runze Yu, Shengli Liu, Yanliang Jin, Shunqing Zhang, Shugong Xu, \textit{Fellow, IEEE}, Xiaoli Chu
\thanks{This work was supported in part by Shanghai Natural Science Foundation under Grant 22ZR1422200, and in part supported by the 6G Science and Technology Innovation and Future Industry Cultivation Special Project of Shanghai Municipal Science and Technology Commission under Grant 24DP1501001. (Corresponding author: Xiaoli Chu)}
\thanks{Yuan Gao, Yiming Liu, Runze Yu, Shengli Liu, Yanliang Jin and Shunqing Zhang are with the School of Communication and Information Engineering, Shanghai University, China, email: gaoyuansie@shu.edu.cn, 1975231949@shu.edu.cn, urleaves@shu.edu.cn, victoryliu@shu.edu.cn, jinyanliang@staff.shu.edu.cn and shunqing@shu.edu.cn.}
\thanks{Shugong Xu is with Xi’an Jiaotong-Liverpool University, Suzhou, China, email: shugong.xu@xjtlu.edu.cn.}
\thanks{Xiaoli Chu is with the Department of Electronic and Electrical Engineering, the University of Sheffield, UK, e-mail: x.chu@sheffield.ac.uk.}}

\maketitle

\begin{abstract}
Fluid antenna systems (FAS) signify a pivotal advancement in 6G communication by enhancing spectral efficiency and robustness. However, obtaining accurate channel state information (CSI) in FAS poses challenges due to its complex physical structure. Traditional methods, such as pilot-based interpolation and compressive sensing, are not only computationally intensive but also lack adaptability. Current extrapolation techniques relying on rigid parametric models do not accommodate the dynamic environment of FAS, while data-driven deep learning approaches demand extensive training and are vulnerable to noise and hardware imperfections. To address these challenges, this paper introduces a novel self-supervised learning network (SSNet) designed for efficient and adaptive channel extrapolation in FAS. We formulate the problem of channel extrapolation in FAS as an image reconstruction task. Here, a limited number of unmasked pixels (representing the known CSI of the selected ports) are used to extrapolate the masked pixels (the CSI of unselected ports). SSNet capitalizes on the intrinsic structure of FAS channels, learning generalized representations from raw CSI data, thus reducing dependency on large labelled datasets. For enhanced feature extraction and noise resilience, we propose a mix-of-expert (MoE) module. In this setup, multiple feedforward neural networks (FFNs) operate in parallel. The outputs of the MoE module are combined using a weighted sum, determined by a gating function that computes the weights of each FFN using a softmax function. Extensive simulations validate the superiority of the proposed model. Results indicate that SSNet significantly outperforms benchmark models, such as AGMAE and long short-term memory (LSTM) networks by using a much smaller labelled dataset. A key observation is that the proposed model is more effectively trained using a small unmasked ratio of known CSI. Specifically, the proposed SSNet trained using CSI of 10 \% total ports outperforms that trained using CSI of 25 \% and 50 \% total ports. This is because using a smaller number of known CSIs during training, the proposed model is forced to learn more effective channel correlation for channel extrapolation at the expense of higher training complexities. Ablation experiments reveal substantial performance gains from the MoE module's integration. Furthermore, zero-shot learning experiments show a moderate performance degradation of about 3-5 dB, underscoring the model's robust generalization ability. Finally, the inference speed experiments illustrate that the proposed model outperforms the benchmark models dramatically at the expense of a slightly longer execution time of 1.13 ms, 2.9 ms, and 3.12 ms on NVIDIA RTX 4090, 4060, and 3060 graphics processing units (GPU)s, respectively.
\end{abstract}


%
\IEEEpeerreviewmaketitle

\section{Introduction}
In the evolution towards the sixth generation mobile communications (6G), fluid antenna systems (FAS) present an innovative paradigm by dynamically adjusting the antenna's form and position to overcome the physical limitations of traditional antennas\cite{wong2022bruce,new2024tutorial}. The core advantage of FAS lies in its ability to control the spatial distribution of fluid radiating elements through software, achieving multidimensional flexibility such as frequency, radiation pattern, and polarization. This enables the system to avoid deep fading and enhance diversity gain in dense multipath environments\cite{faddoul2024advanced}. Research indicates that FAS can significantly increase spectral efficiency, provide new degrees of freedom for massive user access\cite{zou2024shifting,wong2024virtual} and reduce outage probability \cite{xu2024revisiting,new2023fluid,psomas2023diversity}. 

A critical factor in harnessing the potential of FAS is obtaining accurate channel state information (CSI). Traditional CSI acquisition methods, reliant on pilot-based interpolation or compressive sensing, face prohibitive overheads in case of the high-dimensional spatial domain in FAS\cite{xu2023channel,new2024channel,skouroumounis2022fluid,zhang2023successive}. Channel extrapolation techniques \cite{jin2025linformer,zhang2021deep,zhang2023ai} serve as a crucial element in unlocking the potential of FAS by reconstructing complete channel characteristics from partial observation data. Existing research on channel extrapolation can be categorized into two main approaches, and the first relies on parametric models\cite{qiu2024can,liu2022massive,rottenberg2019channel,han2019efficient,jin2023model,zhang2022deep}, such as sparse bayesian learning, which employs sparsity across multiple domains to achieve channel extrapolation\cite{xu2022sparse}. However, this approach depends on prior assumptions about channel structure, making it less adaptable to the highly dynamic nature of FAS. The second approach is data-driven, employing methods like deep neural networks (DNN)\cite{gao2025joint,yu2024multi} to extrapolate CSI\cite{jiang2025towards,gao2025enabling,jin2025linformer}. Nonetheless, the training complexity of these models increases exponentially with antenna array size and they are highly sensitive to hardware imperfections\cite{lin2021deep}.

Due to the importance of acquiring CSI to unleash the potential of FAS, research on channel extrapolation of FAS have been rolled out\cite{waqar2023deep}. \cite{zhang2024learning} proposes an asymmetric graph masked autoencoder (AGMAE) for fluid antenna system channel extrapolation. Simulation results demonstrate that AGMAE achieves accurate CSI reconstruction with low computational complexity, even with a small number of observed antennas, at the ratio of 20\%. A masked language model is proposed for channel extrapolation of FAS and simulation results show that complete CSI can be effectively extrapolated from incomplete observations, improving port selection accuracy and reducing outage probability\cite{wu2024channel}. \cite{wang2024fluid} uses deep reinforcement learning to jointly optimize port selection and precoding in a fluid antenna system for integrated sensing and communication (ISAC)\cite{jin2025dual,gao2024performance,jiang2025c2s,gao2024c2s}. When only partial channel state information is available, a masked autoencoder is used for channel extrapolation, achieving significant sum-rate improvements compared to conventional methods. A model-driven channel extrapolation method for massive fluid antenna systems, leveraging the inherent structure of Jake's rich-scattering channel model is proposed in \cite{li2024model}, and achieves high accuracy with minimal training data and robustly extrapolates CSI for all antenna ports. 

However, research on channel extrapolation of FAS are still in its infancy and the following challenges have not been well addressed. \begin{itemize}
\item Lack of flexibility: The existing methods are not designed to handle a variable number of CSI inputs. They often rely on fixed-size input vectors, making them inflexible and unable to adapt to scenarios where the number of observable ports changes. This limits their applicability to real-world, dynamic environments where the number of available CSI measurements might vary.
\item Large dataset required: The existing models require huge dataset volume to train, for example 800,000 and 120,000 samples are using in training the models in \cite{zhang2024learning} and \cite{wang2024fluid}, respectively. Although such large volume dataset can be relatively easy acquired via simulation, it is inevitable to use measurement CSI data for model training to ensure the applicability of the models in practical scenarios and it is time-consuming and expensive to carry out measurement for such large data volume. 

\item Poor Robustness to noise: The simulation results presented in the existing research focus on normalized mean squared error (NMSE) under noise-free conditions. The absence of experiments with added noise makes their applicability in noisy scenarios in doubt. 
\end{itemize}

Self-supervised learning has emerged as a transformative approach to resolve the above challenges \cite{jin2024hsgan,jin2023sscmt,jin2023multiclass} and have been applied in various fields of wireless communications\cite{jiang2025towards,van2024generative,xu2025enhanced}. A key advantage of self-supervised learning lies in its ability to reduce dependency on labeled datasets\cite{zhang2023self,gui2024survey}, which are costly to collect and often raise privacy concerns. For instance, contrastive learning and autoencoder-based frameworks enable models to learn generalized representations directly from raw CSI data, achieving human activity recognition accuracy comparable to supervised methods much less labeled data\cite{wang2024csi}. \cite{zhang2023self} indicates that self-supervised learning outperforms supervised learning using the same amount of labeled data in channel estimation for a RIS-assisted communication system. Additional, self-supervised learning excels at extracting spatiotemporal and frequency-domain features from high-dimensional CSI. Techniques like channel charting leverage autoencoders to map CSI into low-dimensional latent spaces, preserving spatial correlations for user localization in MIMO systems\cite{ferrand2023wireless}. Channel charting is also exploited in channel prediction by exploiting spatial relationships between known estimates that are embedded in the channel chart\cite{stephan2024channel}. Models trained with self-supervised learning often demonstrate better generalization to unseen data and robustness to noisy or corrupted inputs\cite{dai2023vision,oquab2023dinov2}. Further, by learning underlying patterns and structures rather than memorizing specific input lengths, self-supervised learning can handle variations more effectively\cite{he2022masked}. The ability of self-supervised learning to handle variations in input lengths is crucial for applications involving natural language processing (NLP)\cite{vaswani2017attention}, image processing\cite{rani2023self}, and other high-dimensional data tasks.

\begin{table}[h]
\caption{Mathematical Symbols and Meanings}
\label{tab:math_symbols}
\centering
\begin{tabular}{|c|c|}
\hline
\textbf{Symbol} & \textbf{Meaning}  \\
\hline
$d_{\mathrm{X}}$ & Port spacing in the $x$-direction \\
$d_{\mathrm{y}}$ & Port spacing in the $y$-direction  \\
$W_{\mathrm{X}}$ & Physical dimension of FAS in the $x$-direction \\
$W_{\mathrm{Y}}$ & Physical dimension of FAS in the $y$-direction \\
$N_{x}$ & Number of ports in the $x$-direction \\
$N_{y}$ & Number of ports in the $y$-direction \\
$N_{\mathrm{S}}$ & Total number of ports \\
$\Sigma_{\mathrm{clarke}}(i,j)$ & Spatial correlation matrix of Clarke's model \\
$\mathbf{p}_{i}$ & Position of the $i$-th port \\
$\mathbf{p}_{j}$ & Position of the $j$-th port \\
$\mathbf{U}$ & Matrix of eigenvectors \\
$\mathbf{A}$ & Diagonal matrix of eigenvalues \\
$\mathbf{g}$ & CSI matrix \\
$\delta^{2}$ & Path loss \\
$\mathbf{G}$ & Matrix of i.i.d. complex Gaussian random variables \\
$\mathbf{u}_{\mathrm{b}}$ & CSI of observed ports \\
$\mathbf{W}_{\mathrm{p}}$ & Projection matrix \\
$\mathbf{x}_{\mathrm{p}}$ & Patch embeddings \\
$\mathbf{E}_{pos}$ & Positional embedding matrix \\
$\mathbf{x}^{\prime}_{p}$ & Updated patch embedding sequence \\
$\omega_{k}$ & Frequency vector \\
$\mathbf{Q}$ & Query matrix \\
$\mathbf{K}$ & Key matrix \\
$\mathbf{V}$ & Value matrix \\
$d_{k}$ & Dimension of key vectors \\
$\mathbf{x}_{\rm out}$ & Output of transformer block \\
$\mathbf{z}$ & Latent feature representation from encoder \\
$\mathbf{W}_{d}$ & Projection matrix of decoder\\
$\mathbf{z}^{\prime}$ & Projected latent representation \\
$\mathbf{E}^{dec}_{pos}$ & Positional embedding matrix for decoder \\
$\mathbf{z}^{\prime\prime}$ & Updated latent representation \\
$\mathbf{z}_{out}$ & Output of decoder transformer block \\
$\mathbf{W}_{r}$ & Reconstruction matrix \\
$\mathbf{y}$ & Reconstructed patches \\
$\mathbf{Y}$ & Reconstructed CSI \\
$\mathcal{L}_{\rm CON}$ & Contrastive loss \\
$\mathcal{L}_{\text{NMSE}}$ & Normalized mean squared error loss \\
$\mathcal{L}$ & Combined loss function \\
$\mathcal{E}_i$ & The $i$-th expert network in MoEBlock \\
$\mathcal{G}$ & Gating network in MoEBlock \\
$E$ & Number of experts \\
$K$ & Number of activated experts per token \\
$\mathbf{W}_{e1}^{(j)}$ & First-layer weight matrix of the $j$-th expert \\
$\mathbf{W}_{e2}^{(j)}$ & Second-layer weight matrix of the $j$-th expert \\
$\mathbf{W}_g$ & Weight matrix of the gating network \\
$\mathbf{b}_g$ & Bias vector of the gating network \\
$\boldsymbol{\pi}$ & Gating scores (vector of expert affinities) \\
$s_k$ & Gating weight for the $k$-th selected expert \\
$l_k$ & Index of the $k$-th selected expert \\
$\omega_k$ & Frequency component for positional encoding \\
\hline
\end{tabular}
\end{table}

Inspired by the above research, we proposed a self-supervised learning-based network (SSNet) for channel extrapolation in FASs, tackling the shortcomings of prior work. The key contributions are summarized as follows:
\begin{itemize}

\item We formulate channel extrapolation in FAS as an image reconstruction task. In this framework, a few unmasked pixels (known CSI of selected ports) guide the extrapolation of masked pixels (CSI of unselected ports). Our SSNet leverages the inherent structure of FAS channels to learn generalized representations directly from raw CSI data, thereby minimizing the need for extensive labeled datasets. Our detailed simulations demonstrate that SSNet significantly surpasses benchmark models, such as AGMAE \cite{zhang2024learning} and long short-term memory (LSTM) networks, utilizing just 2.5 \% of the channel dataset.
\item  For enhanced feature extraction and noise resilience, we propose a mix-of-expert (MoE) module. In this setup, multiple feedforward neural networks (FFNs) operate in parallel. The outputs of the MoE module are combined using a weighted sum, determined by a gating function that computes the weights of each FFN using a softmax function. Ablation experiments reveal substantial performance gains from the MoE module's integration. 

\item  A critical finding is that our model could be trained more effectively with a smaller proportion of unmasked known CSI. Specifically, the SSNet trained with CSI from 10 \% of total ports outperforms versions trained with 25 \% and 50 \%. This is because a smaller dataset compels the model to learn more effective channel correlations for extrapolation at the expense of increased training complexity. 

\item Additionally, zero-shot learning experiments indicate a moderate reduction in performance, around 3-5 dB, highlighting the model's robust generalization capabilities. Inference speed tests further reveal that our model achieves significantly better performance over benchmark models, despite slightly longer execution times of 1.13 ms, 2.9 ms, and 3.12 ms on NVIDIA RTX 4090, 4060, and 3060 GPUs, respectively.
\end{itemize}

The structure of this paper is as follows. Section \ref{sec:con} presents the system model for the FAS communication systems and the definition of the channel prediction problem. Section \ref{sec:model} introduces our proposed SSNet model in detail. Section \ref{sec:simulation} details our experimental results using simulated and measurement CSI data with in-depth discussion. Section \ref{sec:system} offers concluding remarks and future directions. Mathematical notations in this manuscript can be found in Table \ref{tab:math_symbols}.

\section{System Model And Channel Extrapolation}
\label{sec:system}
\begin{figure}[htbp]\centering    
\includegraphics[width=1\columnwidth]{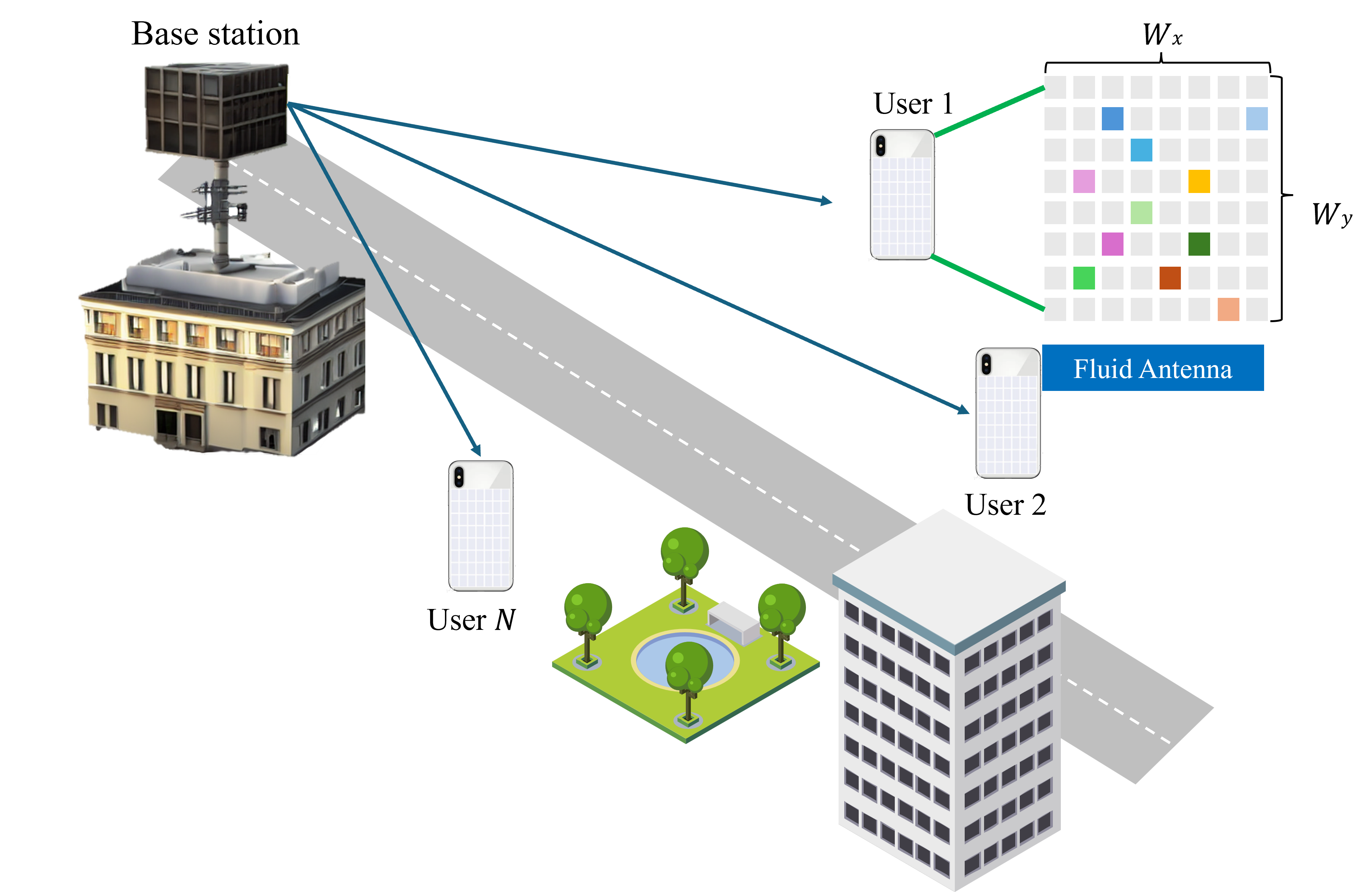} 
\caption{An illustration for communication system with one base station and N users equipped with FAS. }   
\label{fluid_model}\end{figure} 

\begin{figure*}[htbp]\centering    
\includegraphics[width=2\columnwidth]{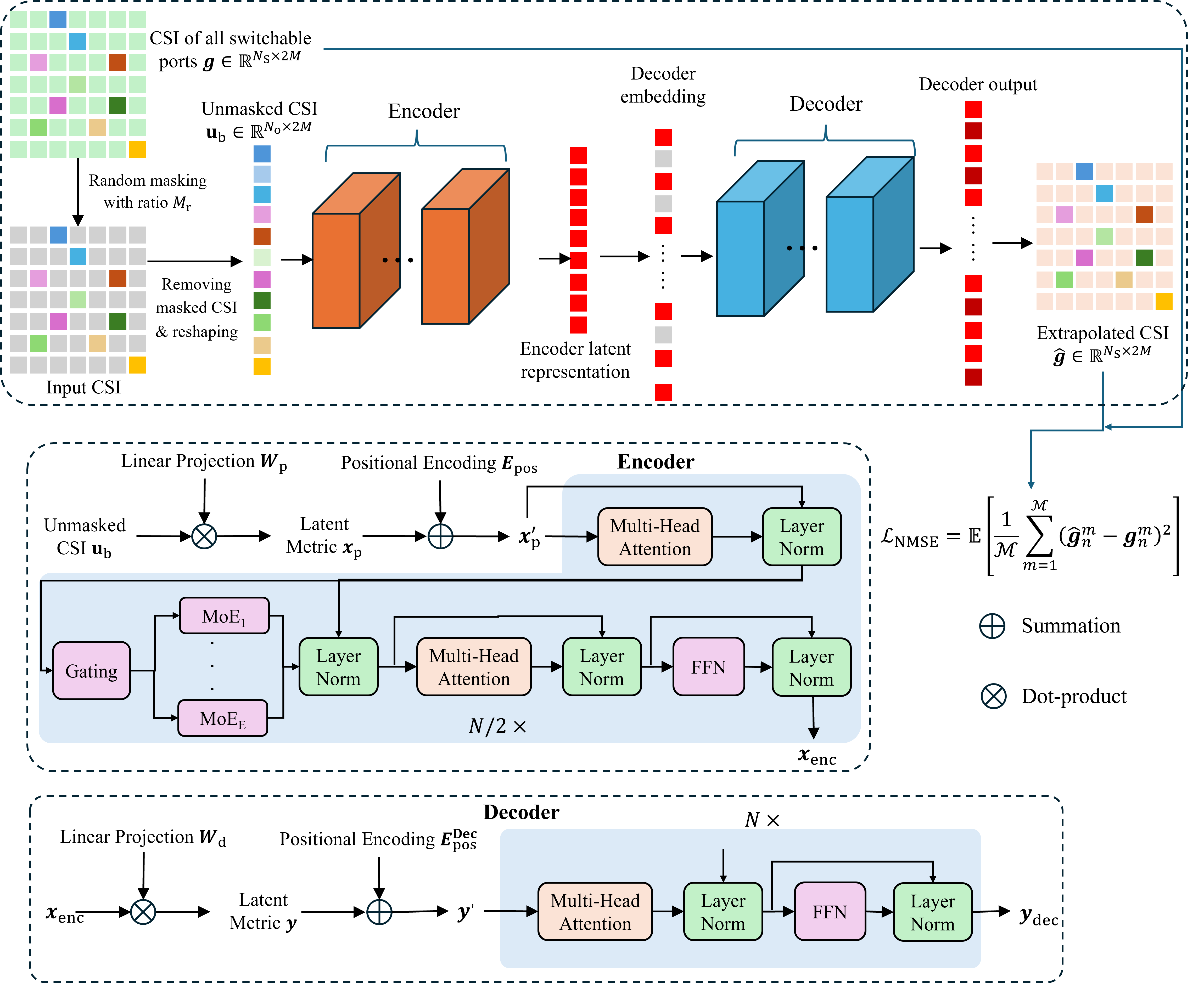} 
\caption{The proposed architecture for CSI extrapolation in FAS.}   
\label{architecture}\end{figure*} 
As illustrated in Fig. \ref{fluid_model}, we consider a communication system with one base station (BS) and $K$ users, where the BS provide communication service for all the $K$ users with the same time-frequency resource. BS implemented $M$ antennas with fixed locations and each user is equipped with an FAS. The FAS is in shape of a 2D plane with a total number of \( N_\text{S}\) switchable ports, which are uniformly distributed in a planar surface with physical size \( W_\text{S}\). Thus, the spacing between ports in the $x$ and $y$ directions is given by:
\begin{equation}\label{toa_aoa_loc}
\left\{\begin{matrix}
 d_\text{X} = \frac{W_\text{X}}{N_\text{X} - 1},\\ 
d_\text{Y} = \frac{W_\text{Y}}{N_\text{Y} - 1},
\end{matrix}\right.
\end{equation}
where \( W_\text{X} \) and \( W_\text{Y} \) are the physical dimensions of the FAS in the $x$ and $y$ directions, respectively, satisfying \( W_\text{S} = W_\text{X} \times W_\text{Y} \). \( N_x \) and \( N_y \) are the number of ports in the $x$ and $y$ directions, respectively, satisfying \( N_\text{S} = N_\text{X} \times N_\text{Y} \).

We exploit Clarke's isotropic scattering model \cite{bjornson2020rayleigh,ramirez2024new} to compute the wireless channel of the FAS and the spatial correlation matrix \( \Sigma_{\text{clarke}} \) for two ports located at positions \( \mathbf{p}_i = (x_i, y_i) \) and \( \mathbf{p}_j = (x_j, y_j) \) is calculated as:
 \begin{equation}\label{eq:clarke_corr}
  \Sigma_{\text{clarke}}(i, j) = \text{sinc}\left(2 \cdot \|\mathbf{p}_i - \mathbf{p}_j\|\right),
\end{equation}
  where \( \text{sinc}(x) = \frac{\sin(\pi x)}{\pi x} \) and \( \|\mathbf{p}_i - \mathbf{p}_j\| \) is the Euclidean distance between the two ports.

The spatial correlation matrix \( \Sigma_{\text{clarke}} \) is decomposed into its eigenvalues and eigenvectors:
\begin{equation}\label{eq:SVG}
  \Sigma_{\text{clarke}} = \mathbf{U} \mathbf{A} \mathbf{U}^H,
\end{equation}
  where \( \mathbf{U} \) is the matrix of eigenvectors, and \( \mathbf{A} \) is the diagonal matrix of eigenvalues. The CSI matrix \( \mathbf{g} \) is generated using the following formula:
\begin{equation}\label{eq:channel_generation}
  \mathbf{g} = \sqrt{\delta^2} \cdot \mathbf{U} \sqrt{\mathbf{A}} \mathbf{G},
\end{equation}
  where \( \delta^2 \) is the path loss, and \( \mathbf{G} \) is a matrix of independent and identically distributed (i.i.d.) complex Gaussian random variables with zero mean and variance \( \frac{1}{2} \).

In the context of FAS, the relationship between the physical location of the antenna ports and the corresponding CSI is inherently bijective. This bijective mapping, denoted as \(\Phi_g : \{\textbf{g}_B\} \rightarrow \{\textbf{g}_A\}\), ensures that each port's position uniquely determines its channel characteristics. The primary objective of channel extrapolation is to infer the CSI of unobserved ports \(\widehat{\textbf{g}}_A\) based on the CSI of observed ports \(\textbf{g}_B\) and minimize the difference between extrapolated CSI $\widehat{\textbf{g}}_A$ and real CSI $\textbf{g}_A$. This can be formally expressed as:

\begin{equation}
     \min_{f(\cdot)}(|\widehat{\textbf{g}}_A-\textbf{g}_A|^2).
\end{equation}where \(f(\cdot)\) represents the mapping operation between the the CSI of observed ports and the CSI of unobserved ports.

\section{Self-supervised learning-based network}
\label{sec:model}
To extrapolate the complete CSI of all the switchable ports using the CSI of a limited number of observable ports, we propose an self-supervised learning-based model for channel extrapolation in FAS. We formulate the channel extrapolation of FAS as an image reconstruction problem, aiming to reconstruct all the pixels of an image-like structure (i.e., the complete CSI of all the switchable ports) using the unmasked pixels (CSI of the observable ports). Our method leverages the principles of MAE, i.e., an encoder-decoder architecture, where the encoder extracts the latent representation of the unmasked CSI, which are subsequently exploited by the decoder to reconstructs the full CSI. The architecture is designed to handle the high-dimensional and non-linear nature of FAS channels, while maintaining computational efficiency and generalization capabilities. 
\subsection{CSI preprocessing}
To model the process that the CSI of only a subset of ports are exploited to extrapolate the CSI of all switchable ports, we first perform masking on the CSI of all ports $\mathbf{g} \in \mathbb{R}^{N_\text{S} \times 2M}$, where $N_\text{S}$ and $M$ are the number of antennas at the BS and of the FAS. $2$ indicates that a CSI contains a real and imaginary part. To mimic the channel acquisition process in practical FAS, the ports with known CSI are randomly selected and the percentage of unknown CSI is denoted as the mask ratio $M_r\in(0,1)$. Due to the limited overhead for channel acquisition in FAS, $M_\text{r}$ is close to 1 in practical systems. Removing the CSI of the masked ports and reshaping the CSI of unmasked ports leads to the CSI of observable ports $\mathbf{u}_b \in \mathbb{R}^{N_\text{o} \times 2M}$, where $N_\text{o}=N_\text{S}\times M_\text{r}$ is the number of observable ports.  
\subsection{Encoder}
$\mathbf{u}_b$ is the input of the encoder, and is first linearly projected into a latent representation $\mathbf{x}_\text{p}$ in high-dimensional latent space as:
   \begin{equation}
   \mathbf{x}_\text{p} =  \mathbf{u}_\text{b}\cdot\mathbf{W}_\text{p} ,
\end{equation}where $\mathbf{W}_\text{p}\in\mathbb{R}^{2M \times d_\text{model}}$ is a learnable projection matrix. \(\mathbf{x}_\text{p} \in \mathbb{R}^{N_\text{o} \times d_\text{model}}\), where $d_\text{model}$ is the embedding dimension. 
   
To incorporate the spatial information of the antenna ports, positional encoding \(\mathbf{E}_{pos}\) added element-wise to the embedding $\mathbf{x}_\text{p}$ as :
   \begin{equation}
   \mathbf{x}_p' = \mathbf{x}_p + \mathbf{E}_{pos}.
   \end{equation}
   
To ensure that $\mathbf{E}_{pos}$ is channel-model-invariant, we adopt the 2D sine-cosine positional encoding that is only determined by the physical location of the port ($i$, $j$), where $i$ and $j$ are the location index of the port in horizontal and vertical axis, respectively. For each antenna port position $(i,j)$ in the FAS grid, we generate a $D$-dimensional positional embedding vector $\mathbf{E}_{pos}(i,j) \in \mathbb{R}^d_\text{model}$, where $\mathbf{E}_{pos}(i,j)$ is calculated as follows:
   
\begin{equation}
        \mathbf{E}_\text{pos}(i,j) = \text{Concat}\left( \mathbf{E}_{\text{row}}(i), \mathbf{E}_{\text{col}}(j) \right),
\end{equation}where $\mathbf{E}_{\text{row}}, \mathbf{E}_{\text{col}}\in \mathbb{R}^{N_\text{o} \times d_\text{model}/2}$. $\mathbf{E}_{\text{row}}$ and $\mathbf{E}_{\text{col}}$ are the positioning encoding in horizontal and vertical axis, respectively.    

\begin{align}
        \mathbf{E}_{\text{row}}(i,2k) &= \sin(i \cdot \omega_k) \\
        \mathbf{E}_{\text{row}}(i,2k+1) &= \cos(i \cdot \omega_k)
    \end{align}
    where $0 \leq k < D/4$. This alternating sine-cosine pattern provides unique representations while maintaining linear relationships between positions. $\omega_k$ is the angular frequency that forms a geometric progression across the embedding dimension:
    \begin{equation}
        \omega_k = \frac{1}{10000^{2k/D}}, \quad 0 \leq k < \lfloor D/4 \rfloor,
    \end{equation}
    where $k$ is the frequency index. This exponential decay schedule ensures wavelengths span multiple orders of magnitude, capturing both fine-grained and coarse spatial relationships.

    For the column index $j$, we compute the remaining $D/2$ dimensions:
    \begin{align}
        \mathbf{E}_{\text{col}}(j,2k) &= \sin(j \cdot \omega_k) \\
        \mathbf{E}_{\text{col}}(j,2k+1) &= \cos(j \cdot \omega_k)
    \end{align}
    using the \textit{same} frequency components $\omega_k$ as the row encoding.

The patch embeddings with positional information are then passed through a series of encoder block. Each encoder block is a s a stack of a multi-head self-attention (MSA) module, an MoE, an MSA and a feed forward network (FFNs) interleaved with residual connections and layer normalization (LayerNorm). The MSA module captures the relationships between different observable antenna ports by computing attention scores:
\begin{equation}
   \text{MSA}(\mathbf{Q}, \mathbf{K}, \mathbf{V}) = \text{Concat}(\text{head}_1,...,\text{head}_h)W^\text{O},\label{Eq:MSA}
\end{equation}where $\mathbf{Q}=\mathbf{x}_p'W^\text{Q}$, \(\mathbf{K}=\mathbf{x}_p'W^\text{K}\), and \(\mathbf{V}=\mathbf{x}_p'W^\text{V}\) are the query, key, and value matrices of the input $\mathbf{x}_p'$, respectively. $W^O\in \mathbb{R}^{hd_v \times d_\text{model}}$ is the learnable metric, where $h$, $d_v$ and $d_\text{model}$ are the number of heads, the dimension of value metric, the dimension of encoder. MSA basically concatenates the attention of all the $h$ heads. $\text{head}_i$ is the attention of the $i$-th
head, which is calculated as:
\begin{equation}
   \text{head}_i=\text{Attention}(\mathbf{x}_p' W_i^\text{Q}, \mathbf{x}_p' W_i^\text{K}, \mathbf{x}_p' W_i^\text{V}),
   \end{equation}where $\text{Attention}(\mathbf{x}_p' W_i^\text{Q}, \mathbf{x}_p' W_i^\text{K}, \mathbf{x}_p' W_i^\text{V})$ is the self-attention, which is calculated as:
\begin{equation}\begin{aligned}
   &\text{Attention}(\mathbf{x}_p' W_i^\text{Q}, \mathbf{x}_p' W_i^\text{K}, \mathbf{x}_p' W_i^\text{V}) \\&= \text{softmax}\left(\frac{\mathbf{x}_p'W_i^\text{Q}\mathbf{x}_p'W_i^\text{K}}{\sqrt{d_k}}\right)\mathbf{x}_p'W_i^\text{V},
\end{aligned}\end{equation}where $W_i^\text{Q}\in \mathbb{R}^{d_\text{model}\times d_k }$, $W_i^\text{K}\in \mathbb{R}^{d_\text{model}\times d_k }$, and $W_i^\text{V}\in \mathbb{R}^{d_\text{model}\times d_v }$ are learnable projection metrics of the query, key, and value for the $i$-th head, respectively. A residual connection if performed, then follows a LayerNorm as: 
   \begin{equation}
   \mathbf{x}_p'' = (\text{LayerNorm}(\mathbf{x}_p' + \text{MSA}( \mathbf{x}_p'))).
 \end{equation}

To further enhance the feature extraction capabilities and noise robustness of the SSNet framework, we integrate MoE layers subsequently. As illustrated in Fig. \ref{architecture}, the MoE comprises $E$ expert networks $\{\mathcal{E}_1,...,\mathcal{E}_E\}$ and a trainable gating network $\mathcal{G}$. Each expert implements a nonlinear transformation:
\begin{equation}
\mathcal{E}_j({x}_p'') =\sigma\left(\mathbf{x}_p''\mathbf{W}_{e1}^{(j)}  \right) \mathbf{W}_{e2}^{(j)} 
\label{eq:expert}
\end{equation}
where $\sigma$ denotes the GELU activation function \cite{lee2023mathematical}, with weight matrices $\mathbf{W}_{e1}^{(j)} \in \mathbb{R}^{D \times D_h}$, $\mathbf{W}_{e2}^{(j)} \in \mathbb{R}^{D_h \times D}$ and $D_h = 4D$ following the hidden dimension expansion convention in Transformers \cite{vaswani2017attention}.

The gating network computes expert selection probabilities using a softmax-activated linear transformation:
\begin{equation}
\mathbf{g} = \text{softmax}\left({x}_p'' \mathbf{W}_g + \mathbf{b}_g\right), 
\label{eq:gating}
\end{equation}where $\mathbf{W}_g \in \mathbb{R}^{D \times E}$ is the learnable metric for gating. To maintain computational efficiency, we employ top-$K$ sparse activation:
\begin{equation}
\{s_k\}_{k=1}^{K} = \text{TopK}(\mathbf{g}, K)
\label{eq:topk}
\end{equation}
where only the top-$K$ experts process each input token, which preserves the capability of the model and effectively reduces the computational complexity compared to dense activation.

The final output aggregates weighted expert contributions with dropout regularization of probability $p$:
\begin{equation}
\mathbf{x}_p''' = \text{Dropout}\left(\sum_{k=1}^{K} s_k \cdot \mathcal{E}_{l_k}(\mathbf{x})\right).
\label{eq:aggregation}
\end{equation} 

$\mathbf{x}_p'''$ is further computed by a stack of MSA and FFN interleaved with residual connections and LayerNorm as:
\begin{equation}
\mathbf{x}_p'''''= \text{LayerNorm}(\mathbf{\mathbf{x}_p''''}+\text{FFN}(\mathbf{\mathbf{x}_p''''})),
\label{eq:FFN}
\end{equation}where 
\begin{equation}
\mathbf{x}_p''''= \text{LayerNorm}(\mathbf{\mathbf{x}_p'''}+\text{MSA}(\mathbf{\mathbf{x}_p'''})),
\label{MSA_again}
\end{equation} and $\text{FFN}(\cdot)$ is the computing of the multilayer perceptron (MLP) \cite{pinkus1999approximation}.

$\mathbf{x}_p'$ is computed through Eq. (\ref{Eq:MSA}) (\ref{eq:FFN}) for $N/2$ times, which results in the output of the encoder $x_\text{enc}$.

\subsection{Decoder}

The decoder is designed to be lightweight, with fewer layers and a smaller hidden dimension compared to the encoder. It takes the latent representation from the encoder and appends mask tokens to reconstruct the full CSI. The decoder uses a lightweight vision Transformer (ViT) to exploit the local correlations and smoothness of the FAS channels. This local diffusion mechanism allows the decoder to efficiently recover the CSI of the masked ports by propagating information from the observable ports. The latent representation $\mathbf{x}_\text{enc}$ from the encoder is first projected into a lower-dimensional space using a linear transformation:
   \begin{equation}
   \mathbf{y} = \mathbf{x}_\text{enc}\mathbf{W}_d,
   \end{equation}where \(\mathbf{W}_d\) is a learnable projection matrix, and \(\mathbf{z}' \in \mathbb{R}^{N_\text{o} \times D_{decoder}}\) is the projected latent representation with reduced dimensionality \(D_{decoder}\). 
   
   Same as the positioning encoding of the encoder, channel-invariant 2D sine-cosine positional encoding \(\mathbf{E}_\text{pos}^\text{dec}\) are added to the projected latent representation $\mathbf{y}$ to incorporate spatial information:
   \begin{equation}
   \mathbf{y}' = \mathbf{y} + \mathbf{E}_\text{pos}^\text{dec}.
   \end{equation}
   
The decoder applies a series of lightweight transformer blocks to process the latent representation. Each block is a stack of an MSA module, an FFN interleaved with residual connections and LayerNorm. The computing of each Transformer for a given $\textbf{z}$ is given as:
\begin{equation}
\mathbf{z}''= \text{LayerNorm}(\mathbf{\mathbf{z}'}+\text{FFN}(\mathbf{\mathbf{z}'})),
\label{eq:FFN_decoder}
\end{equation}where $\mathbf{z}'$ is calculated as:
   \begin{equation}
   \mathbf{z}' = \text{LayerNorm}(\mathbf{z}+\text{MSA}(\mathbf{z})),\label{Eq:MSA_decoder}
   \end{equation}where $\text{MSA}(\cdot)$ is given in Eq. (\ref{Eq:MSA}). $ \mathbf{y}'$ is computed by Eq. (\ref{eq:FFN_decoder}) and (\ref{Eq:MSA_decoder}) for $N$ times, which results in the output of decoder $y_\text{dec}$. 

\subsection{Output}
   The output of the decoder is $y_\text{dec}$ passed through a reconstruction head, which maps the latent representation back to the original pixel space. This is achieved using a linear projection:
   \begin{equation}
   \mathbf{z} = y_\text{dec}\mathbf{W}_r ,
   \end{equation}
   where \(\mathbf{W}_r\) is the reconstruction matrix, and \(\mathbf{z} \in \mathbb{R}^{N_\text{S} \times 2M}\) represents the reconstructed patches. 
   
Finally, the reconstructed patches are reshaped back into the original CSI format using the unpatchify operation. This involves rearranging the patch sequence into 2-dimensional:
\begin{equation}
\widehat{\mathbf{g}} = \text{Reshape}(\mathbf{y}),
\end{equation}.


\section{Simulation}
\label{sec:simulation}
\subsection{Simulation Settings}

In our simulations, we consider a downlink system where a base station equipped with \( M = 8 \) fixed-position antennas communicates with \( K = 8 \) userss, each employing a two-dimensional FAS. The FAS has a physical size of \( W_\text{s}\) and is uniformly distributed with \( N_\text{s} = 16 \times 32 \) switchable ports. The operating frequency is 3.5 GHz, corresponding to a wavelength of \( \lambda = \frac{c}{f} \), where \( c \) is the speed of light. The path loss is characterized by \( \delta^2 = 1 \). The settings for simulation can be found in Table \ref{tab:simulation_settings}.

\begin{table}[htbp]
\centering\caption{Simulation Settings}
\label{tab:simulation_settings}
\begin{tabular}{cc}
\hline
\textbf{Parameter} & \textbf{Settings} \\ \hline
BS Antennas (\( M \)) & 8 \\ 
Antenna System Type & 2D Fluid Antenna System (FAS) \\ 
FAS Physical Size (\( W_\text{s} \)) & \(8 \times 16\) \& \(2 \times 4 \text{cm}^2\) \\ 
Switchable Ports (\( N_s \)) & 16 \(\times\) 32 (512 in total) \\ 
Operating Frequency (\( f \)) & 3.5 GHz \\ 
Wavelength (\( \lambda \)) & $8.57 \text{cm}$ \\ 
Path Loss (\( \delta^2 \)) & 1 \\ 
Signal to noise ratio & \(0, 10, 20 \)dB \\ \hline
\end{tabular}
\end{table}

The dataset consists of 20,000 noise-free channel samples, which is divided into training and testing sets with an 80:20 ratio. The quality of the channel extrapolation is evaluated using the NMSE calculated as follows: 

\begin{equation}\label{NMSE}
\mathcal{L}_{\text{NMSE}} = \mathbb{E}\left[\frac{1}{\mathcal{M}\mathcal{N}}\sum_{m=1}^{\mathcal{N}}\sum_{n=1}^{\mathcal{M}}(\mathbf{\widehat{g}}_{n}^{m} - \mathbf{g}_{n}^{m})^{2}\right],
\end{equation}where \(\mathcal{M}\) and \(\mathcal{N}\) represent the number of masked ports utilized in each sample and the number of samples tested, respectively. \(\mathbf{\widehat{g}}_{n}^{m}\) represents the predicted CSI for the \(m\)-th masked port of the \(n\)-th sample, \(\mathbf{g}_{n}^{m}\) represents the true masked CSI. The NMSE is normalized by the energy of the target CSI, ensuring that the it is scale-invariant and focuses on the relative error. 

For comparison, we adopt AGMAE as the benchmark model and use 800,000 noise-free channel samples for training, which is same as that in \cite{zhang2024learning}. Another benchmark is the long short-term memory (LSTM), which has been widely used in series prediction tasks, such as time-domain channel prediction\cite{jin2025linformer}. LSTM is tailored to predict the masked CSI using the known CSI.

\begin{figure}[htbp]\centering    
\includegraphics[width=1.05\columnwidth]{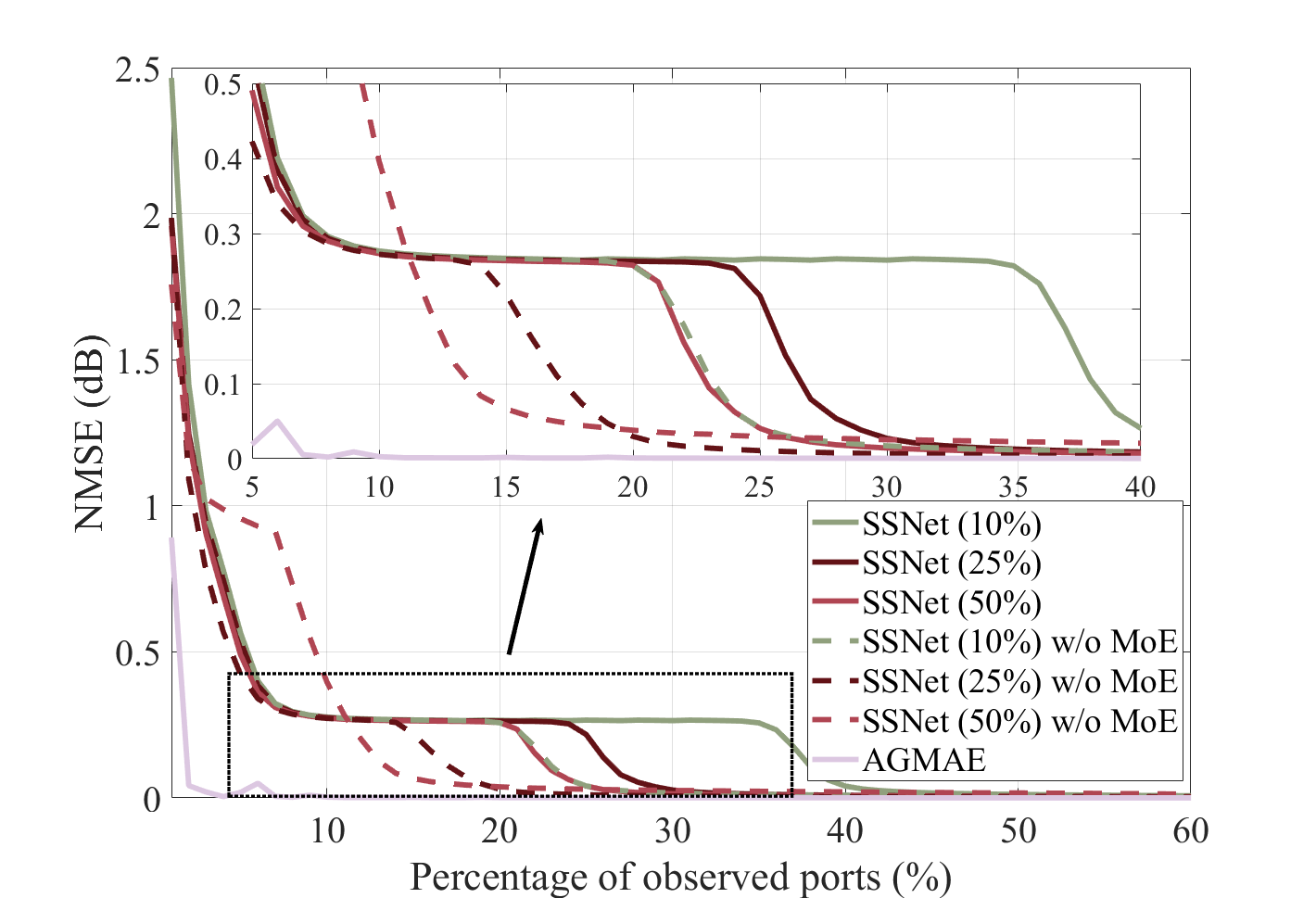} 
\caption{An illustration for training process of SSNet trained with various mask ratio.}  
\label{comparison_loss_[50,75,90]_2_4}\end{figure}

\begin{table}    \caption{Pre-training setting.}
    \label{tab:pre_training_setting}
    \centering
    \begin{tabular}{cc}
     \hline   Training Settings                & Settings                  \\\hline
        Optimizer             & AdamW             \\
        Base learning rate    & 1.5e-4                 \\
        Weight decay          & 0.05                   \\
        Optimizer momentum    & \(\beta_1, \beta_2=0.9, 0.98\) \\
        Batch size            & 64                   \\
        Learning rate schedule& cosine decay      \\
        Warmup epochs         & 40                     \\
        Augmentation          & Random Resized Crop   \\   
        Dropout probability & 0.1\\
        Number of total experts & 4\\
        Number of activated experts &2\\
        Training hardware platform & NVIDIA RTX 4090\\
        Testing hardware platform & NVIDIA RTX 4090, 4060 and 3060
        \\\hline
    \end{tabular}
\end{table}
\begin{figure*}[!] \centering 
\subfigure[The Original CSI.] { 
\label{Imag_20dB_Original_CSI}     
\includegraphics[width=0.47\columnwidth]{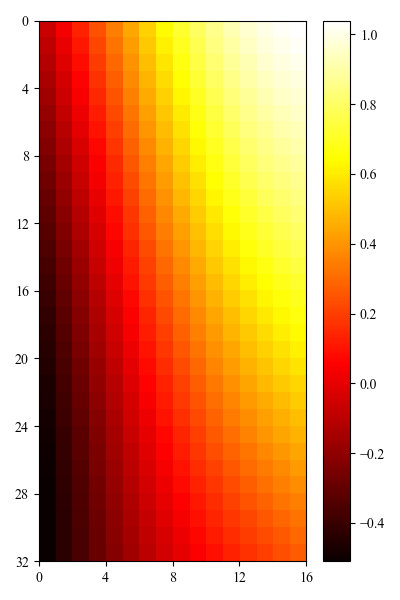} 
}  
\subfigure[Output of SSNet (10\%).] { 
\label{Output_of_SSNetMoe(10)_at_0db}     
\includegraphics[width=0.47\columnwidth]{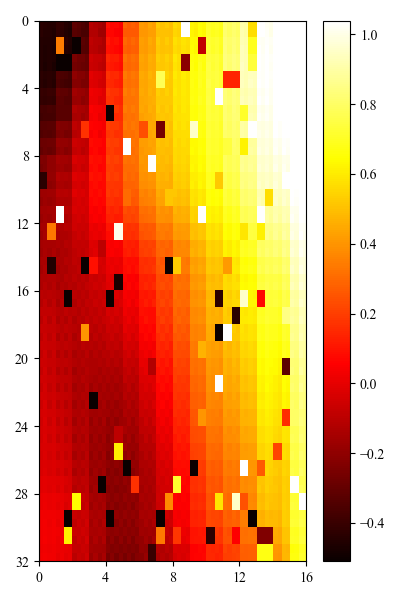} 
} 
\subfigure[Output of SSNet (25\%).] { 
\label{Output_of_SSNetMoe(25)_at_0db}     
\includegraphics[width=0.47\columnwidth]{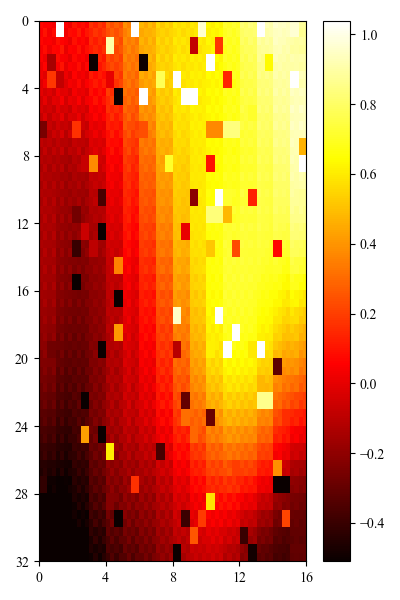} 
}    
\subfigure[Output of AGMAE\cite{zhang2024learning}.] { 
\label{Imag_0dB_mask0.9_Output_CSI_com}     
\includegraphics[width=0.47\columnwidth]{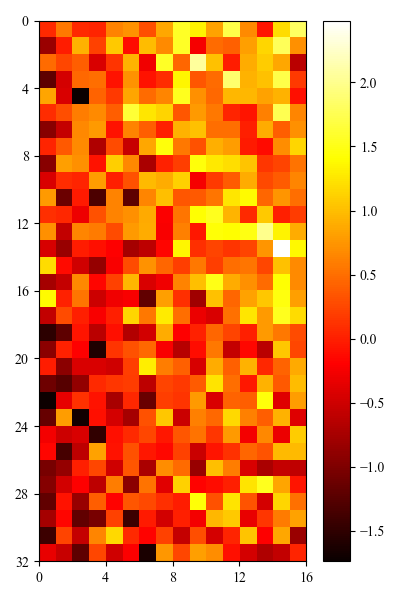} 
}  
\caption{Illustration of the real part of the channel extrapolation for $2\times 4$ cm with $10$ \% observed ports at 0 dB. As the Clarke channel is modeled as complex Gaussian and symmetric, the simulation results of the imaginary part convey essentially the same information as the real part, and are omitted for conciseness.}   
\label{CSI_data_imag_2_4_0dB}   
\end{figure*}

\begin{figure*}[!] \centering 
\subfigure[The Original CSI.] { 
\label{Imag_20dB_Original_CSI}     
\includegraphics[width=0.47\columnwidth]{Imag_Original_CSI.png} 
}  
\subfigure[Output of SSNet (10\%).] { 
\label{Output_of_SSNetMoe(10)_at_20db}     
\includegraphics[width=0.47\columnwidth]{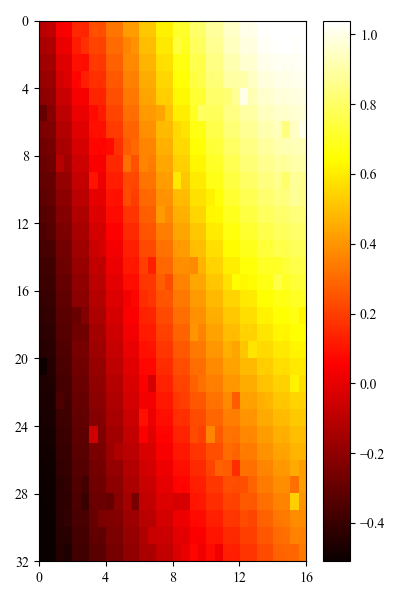} 
} 
\subfigure[Output of SSNet (25\%).] { 
\label{Output_of_SSNetMoe(25)_at_20db}     
\includegraphics[width=0.47\columnwidth]{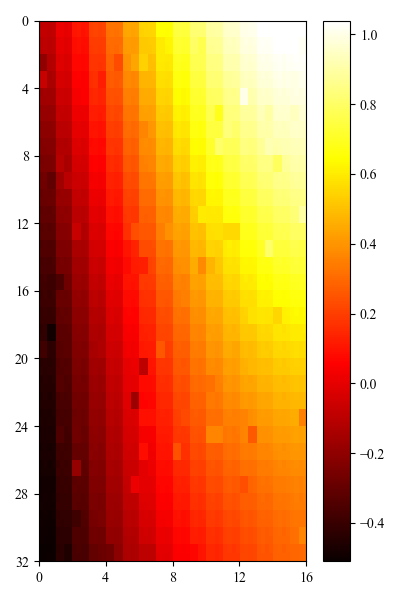} 
}    
\subfigure[Output of AGMAE\cite{zhang2024learning}.] { 
\label{Imag_20dB_mask0.9_Output_CSI_com}     
\includegraphics[width=0.47\columnwidth]{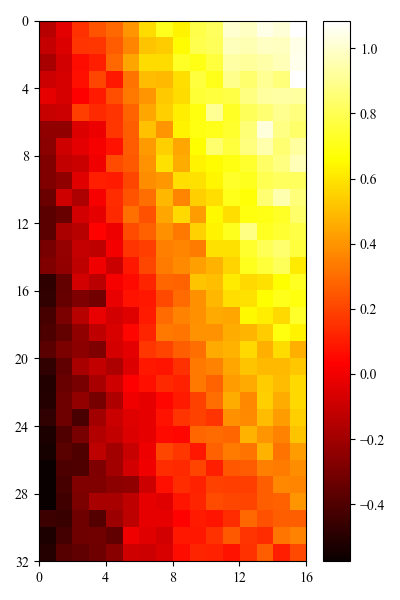} 
}  
\caption{Illustration of the real part of the channel extrapolation for $2\times 4$ cm with $10$ \% observed ports at 20 dB. As the Clarke channel is modeled as complex Gaussian and symmetric, the simulation results of the imaginary part convey essentially the same information as the real part, and are omitted for conciseness.}   
\label{CSI_data_imag_2_4_20dB}   
\end{figure*}
\subsection{Model Training}
The proposed model is traning in a self-supervised manner and the training configurations are summarized in Table \ref{tab:pre_training_setting}. Specifically, we employed the AdamW optimizer, configured with an initial learning rate of \(1.5 \times 10^{-4}\), a weight decay of 0.05, and a cosine decay schedule for the learning rate. Techniques like color jittering, drop path, or gradient clipping are excluded to maintain simplicity. All Transformer blocks are initialized via the Xavier weight initialization method, with a batch size of 64 and a 40-epoch warm-up period included in the training process. SSNet is rained using fixed masking ratios of 75\% and 90\% (denoted as SSNet (25\%) and SSNet (10\%) in the following context, respectively) on FAS of $2\times4$ cm and $8\times16$ cm to ensure the model adapts to a range of observation conditions. During evaluation, we test the model on datasets with varying masking ratios to confirm its ability to generalize. Experiments are performed using PyTorch 2.0.1 and DGL 1.1.2, with computations accelerated on an NVIDIA GeForce RTX 4090 GPU. 

Fig. \ref{comparison_loss_[50,75,90]_2_4} illustrates the change of loss for SSNet (10\%) and  SSNet (25\%) during training, which demonstrate a consistent decline, starting from an initial value of approximately 2. Over the 40 warm-up epochs, the loss decreases at distinct rates, settling at the order of \(10^{-2}\). Beyond this phase, the loss decreases and eventually stabilize at \(10^{-4}\). Notably, the loss of SSNet (10\%) keeps approximately constant at the order of \(10^{-2}\) for over 30 epochs, which is much larger than that of SSNet (25\%), indicating that it is more challenging to train SSNet with a smaller percentage of observed ports.

\subsection{Results Analysis}

\subsubsection{Channel extrapolation accuracy}
The extrapolation accuracy of the proposed SSNet is comprehensively evaluated against AGMAE\cite{zhang2024learning} under varying channel environments and mask ratios. 

We first compare the performance of SSNet and AGMAE quantitatively. Fig. \ref{CSI_data_imag_2_4_0dB} and \ref{CSI_data_imag_2_4_20dB} demonstrate the real part of the channel extrapolation for $2\times 4$ cm FAS with SNR of 0 and 20 dB with 10\% observed ports. Specifically, the comparison between Fig. \ref{Output_of_SSNetMoe(10)_at_0db}-\ref{Imag_0dB_mask0.9_Output_CSI_com} and \ref{Output_of_SSNetMoe(10)_at_20db}-\ref{Imag_20dB_mask0.9_Output_CSI_com} indicates that it is more challenging to extrapolate FAS channel in 0 dB than 20 dB scenarios, which is intuitive. In addition, Fig. \ref{Output_of_SSNetMoe(10)_at_20db}-\ref{Imag_20dB_mask0.9_Output_CSI_com} demonstrate that SSNet (10\%) and SSNet (25\%) outperform AGAME at 20 dB scenarios. The performance gap between AGAME and SSNet (either SSNet (10\%) and SSNet (25\%)) becomes more significant at 0 dB scenarios, as illustrated in Fig. \ref{Output_of_SSNetMoe(10)_at_0db}-\ref{Imag_0dB_mask0.9_Output_CSI_com}. This indicates that SSNet shows high robustness to noise, while AGMAE works poorly at 0 dB scenarios. As the Clarke channel is modeled as complex Gaussian and symmetric, the simulation results of the imaginary part convey essentially the same information as the real part, and are omitted for conciseness.

\begin{figure}[!] \centering 
\subfigure[0 dB.] { 
\label{comparison_4_[0]_[5,10,15,20]_2_4}     
\includegraphics[width=1.05\columnwidth]{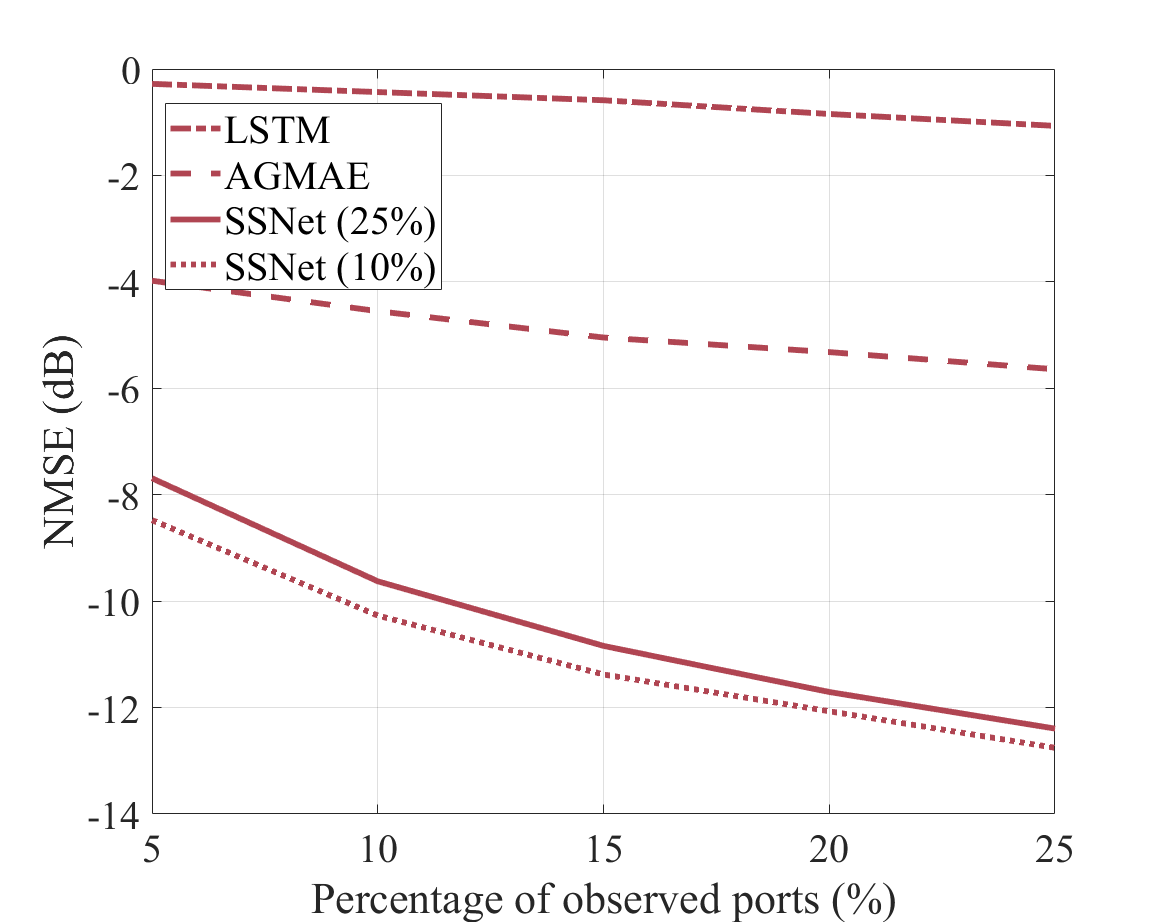} 
}    
\subfigure[20 dB.] { 
\label{comparison_4_[20]_[5,10,15,20]_2_4}     
\includegraphics[width=1.05\columnwidth]{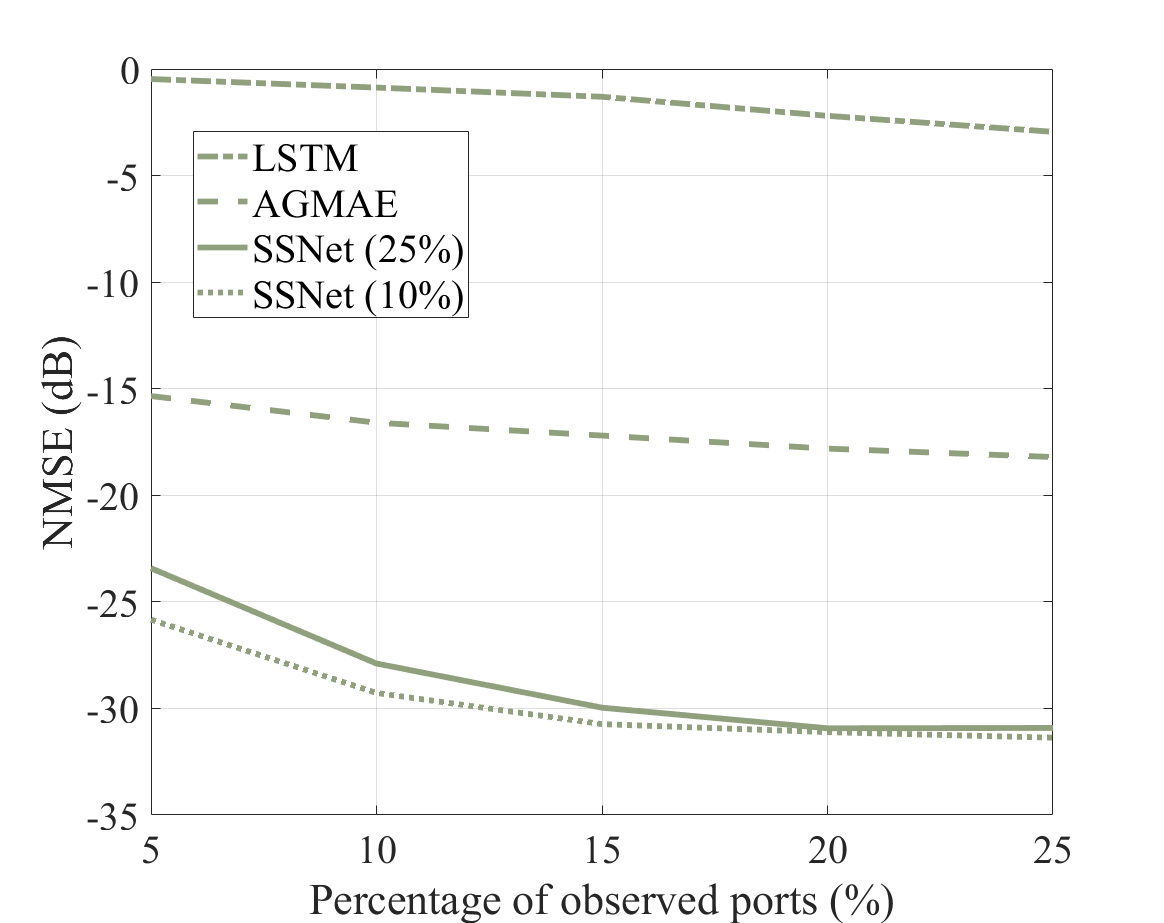} 
}  
\caption{Comparison of SSNet, LSTM and AGMAE\cite{zhang2024learning} in for FAS with of $2\times 4$ cm.}   
\label{comparison_4_space_2_4}   
\end{figure}

\begin{figure}[!] \centering 
\subfigure[0 dB.] { 
\label{comparison_4_[0]_[5,10,15,20]_8_16}     
\includegraphics[width=1.05\columnwidth]{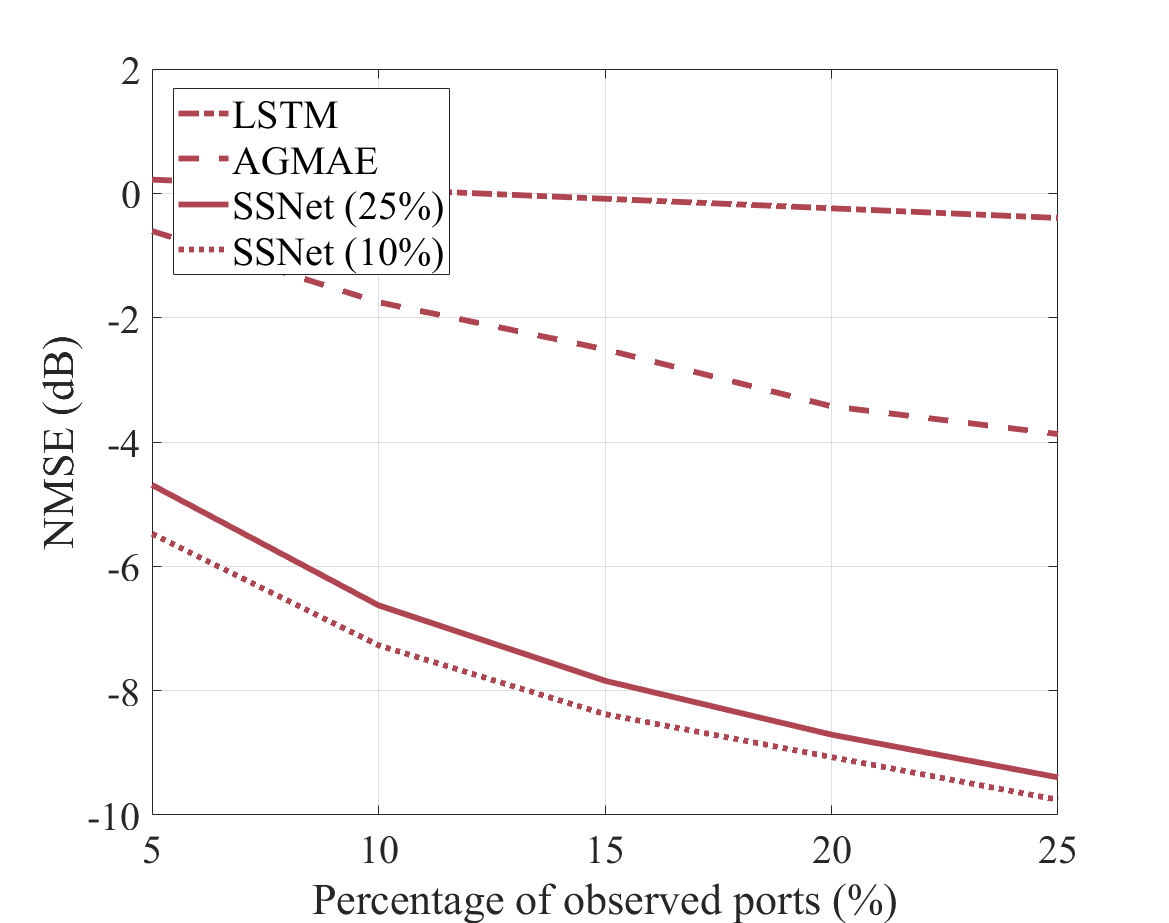} 
}    
\subfigure[20 dB.] { 
\label{comparison_4_[20]_[5,10,15,20]_8_16}     
\includegraphics[width=1.05\columnwidth]{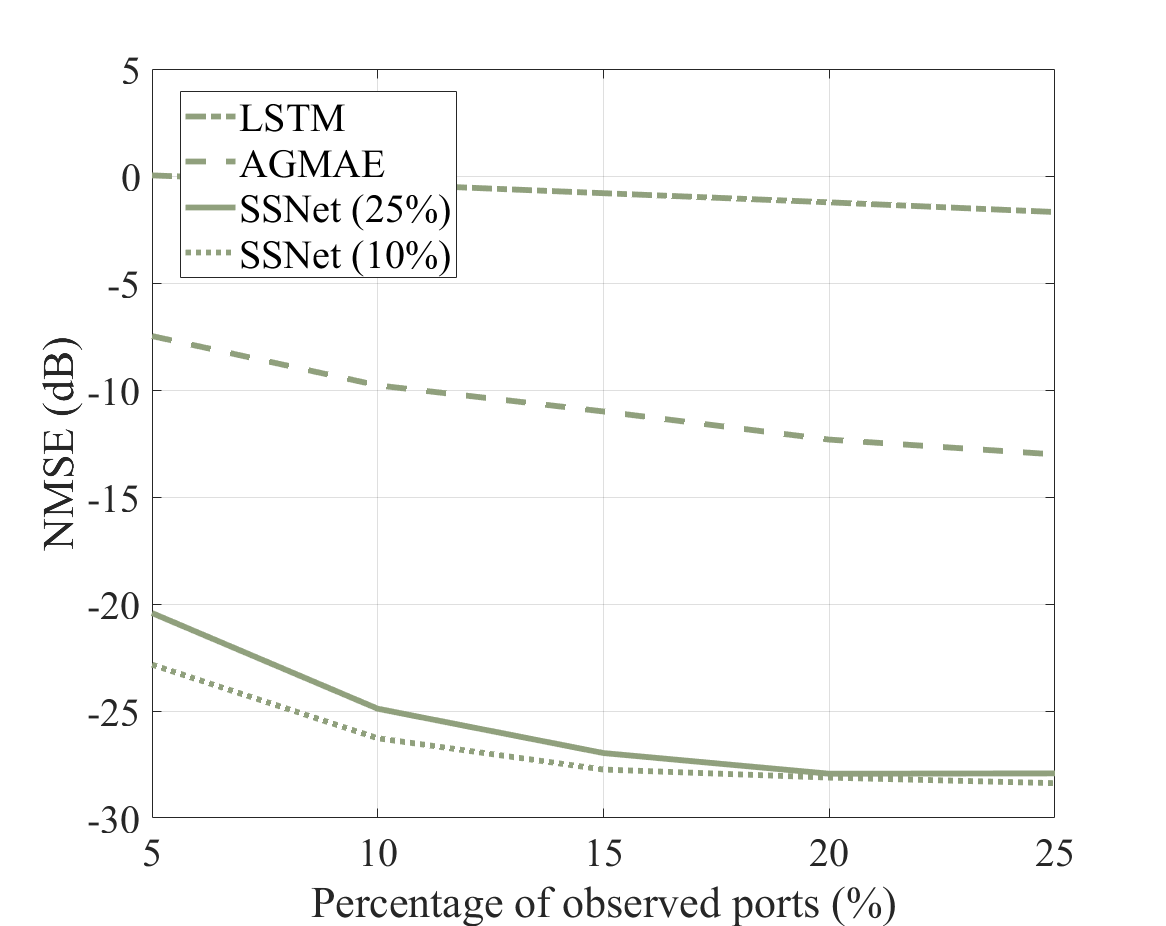} 
}  
\caption{Comparison of SSNet, LSTM and AGMAE\cite{zhang2024learning} in for FAS with of $8\times 16$ cm.}   
\label{comparison_4_space_8_16}   
\end{figure}

To further support our quantitative observations, we carried out quantitative comparison in Fig. \ref{comparison_4_space_2_4} and \ref{comparison_4_space_8_16}, where both SSNet (25\%) and SSNet (10\%) significantly outperform AGMAE and LSTM with various percentage of observed ports. For instance in Fig. \ref{comparison_4_[0]_[5,10,15,20]_2_4}, at 10\% observed ports with SNR of 0 dB, SSNet (10\%), SSNet (25\%), AGMAE and LSTM achieve NMSE of around $-10.2$ dB, $-9.1$ dB, $-4.7$ dB and $-0.4$ dB, respectively. SSNet (10\%) and SSNet (25\%) outperform AGMAE by around 117\% and 93\%, respectively. Illustrated in Fig. \ref{comparison_4_[0]_[5,10,15,20]_2_4}, at 10\% observed ports with SNR of 20 dB, SSNet (10\%), SSNet (25\%) AGMAE and LSTM achieve NMSE of around $-29.3$ dB, $-26.1$ dB and $-16.6$ dB, respectively. SSNet (10\%) and SSNet(25\%) outperform AGMAE by around 76\% and 56\%, respectively. This echos with the observation that the performance gap between AGAME and SSNet (either SSNet (10\%) and SSNet (25\%)) becomes more significant at 0 dB scenarios than 20 dB scenarios.

Fig. \ref{comparison_4_space_8_16} illustrates the channel extrapolation results for $8\times 16$ cm FAS. However, the NMSE in $8\times 16$ cm FAS is larger than that in $2\times 4$ for any given model, SNR and percentage of observed ports. Taking 10\% observed ports with SNR of 0 dB case in Fig. \ref{comparison_4_[0]_[5,10,15,20]_8_16} as an example, the NMSE of SSNet (10\%), SSNet (25\%), AGAME and LSTM are $-7.3$ dB, $-6.6$ dB, $-1.7$ dB and $0.6$ dB for $8\times 16$ cm FAS, while the corresponding NMSE for $2\times 4$ cm FAS are $-10.2$ dB, $-9.1$ dB, $-4.7$ dB and $-0.4$ dB, respectively. This indicates that channel extrapolation in $8\times 16$ cm is more challenging than that in $2\times 4$ cm, which is attributed that the correlation between adjacent antenna is more significant in FAS of a smaller size.

\begin{figure}[!] \centering 
\subfigure[5 \% of ports are observable.] { 
\label{comparison_2_[0,10,20]_[5]_2_4_8_16}     
\includegraphics[width=1.05\columnwidth]{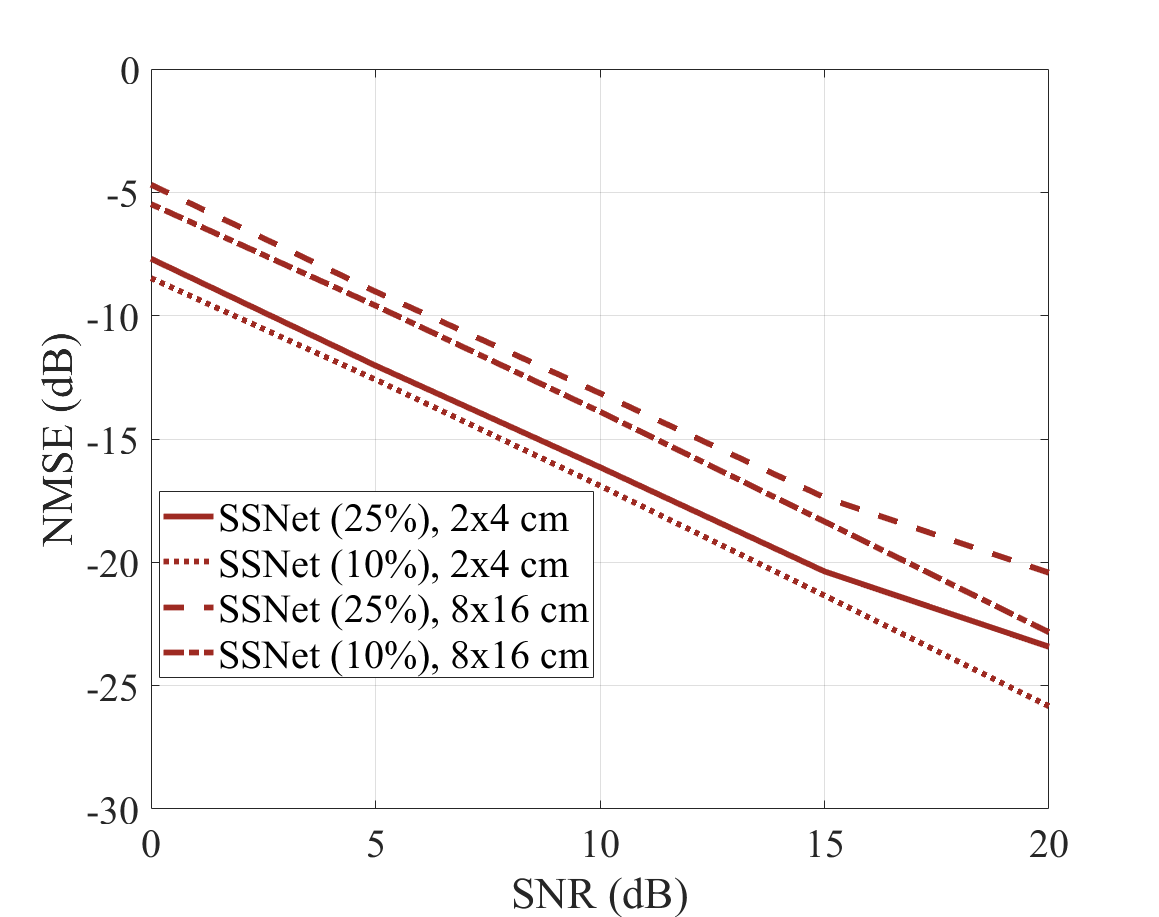} }    
\subfigure[15 \% of ports are observable.] { 
\label{comparison_2_[0,10,20]_[15]_2_4_8_16}     
\includegraphics[width=1.05\columnwidth]{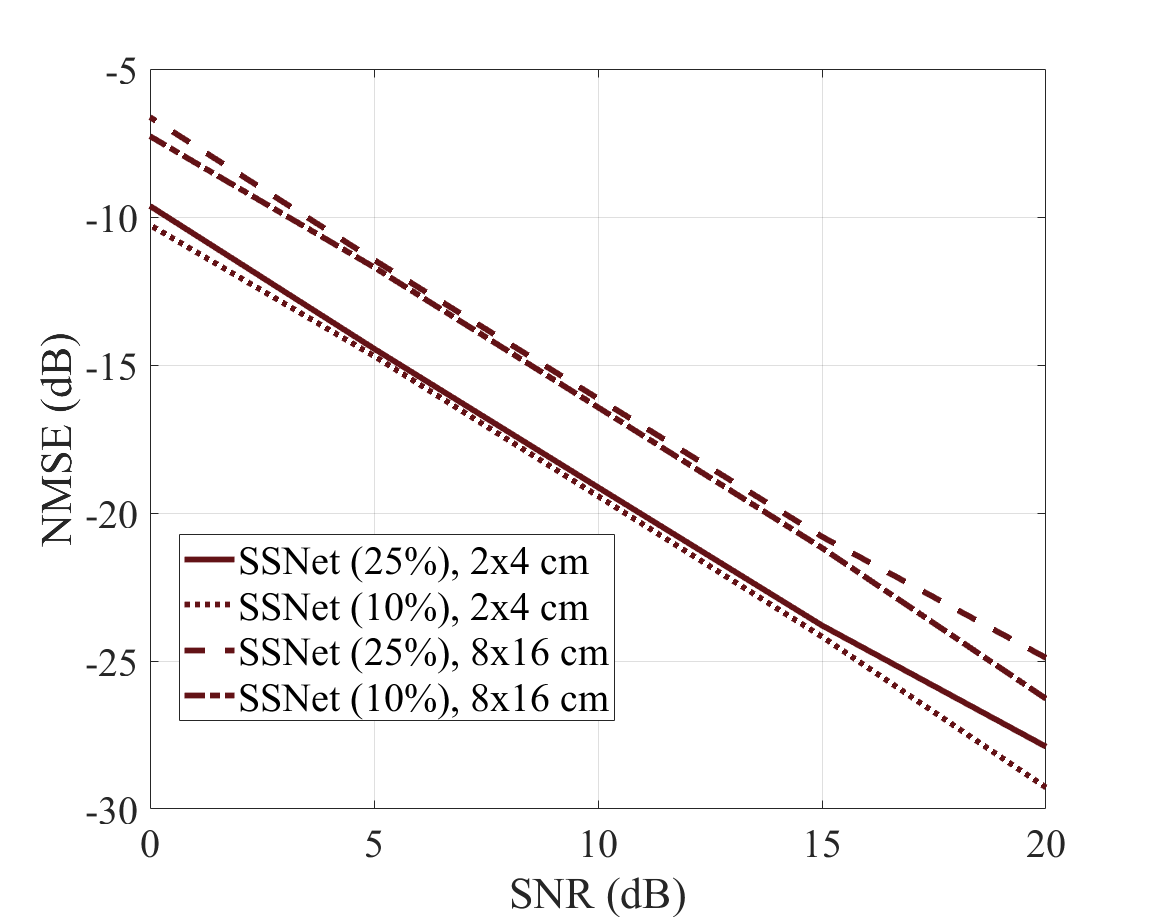} 
}  
\caption{An illustration for the channel extrapolation accuracy of SSNet with various SNR levels. (SSNet $(x\%), y\times z$) indicates the case that SSNet (x\%) is tested in $y\times z$ cm FAS with $s$\% of observed ports.} 
\label{MR_SNR}\end{figure}

As illustrated in Fig. \ref{MR_SNR}, intuitively, the channel extrapolation performance of SSNet (10\%) and SSNet (25\%) degrade with the increase of noise level. Notably, as illustrated in Fig. \ref{comparison_2_[0,10,20]_[5]_2_4_8_16} abd \ref{comparison_2_[0,10,20]_[15]_2_4_8_16}, the performance gap between SSNet (10\%) and SSNet (25\%) with various percentages of observed ports for $2\times 4$ cm FAS is larger than that for $8\times 16$ cm FAS. This indicates that SSNet (10\%) is more effective in learning the correlation between antennas than SSNet (10\%) for $2\times 4$ cm FAS.

Last but not least, both SSNet (25\%) and SSNet (10\%) significantly outperform AGMAE using much less training sample. Specifically, we use 20,000 CSI samples to train SSNet (25\%) and SSNet (10\%), and 800,000 CSI samples to train AGAME. This indicates the SSNet outperforms AGAME with only 2.5\% of training samples, demonstrating that SSNet is significantly more data-efficient than AGAME.

\subsubsection{Effect of mask ratio in training SSNet}
The impact of mask ratios for training SSNet is also revealed. We first compare SSNet (25\%) and SSNet (10\%) qualitatively. Generally, SSNet (10\%) outperforms SSNet (25\%) in terms of real and imaginary part of the channel extrapolation for 2 × 4 cm FAS with SNR of 0 and 20 dB with 10\% observed ports. Specifically, comparing Fig. \ref{Output_of_SSNetMoe(10)_at_20db} and \ref{Output_of_SSNetMoe(25)_at_20db}, we observe that SSNet (10\%) outperforms SSNet (25\%) in 20 dB scenarios. The same trend is revealed in 0 dB scenarios, as illustrated in Fig. \ref{Output_of_SSNetMoe(10)_at_0db} and \ref{Output_of_SSNetMoe(25)_at_0db}.

\begin{figure}[!] \centering 
\subfigure[0 dB.] { 
\label{ablation_comparison_2_[0db]_[5,10,15,20,25,50]_2_4}
\includegraphics[width=1.\columnwidth]{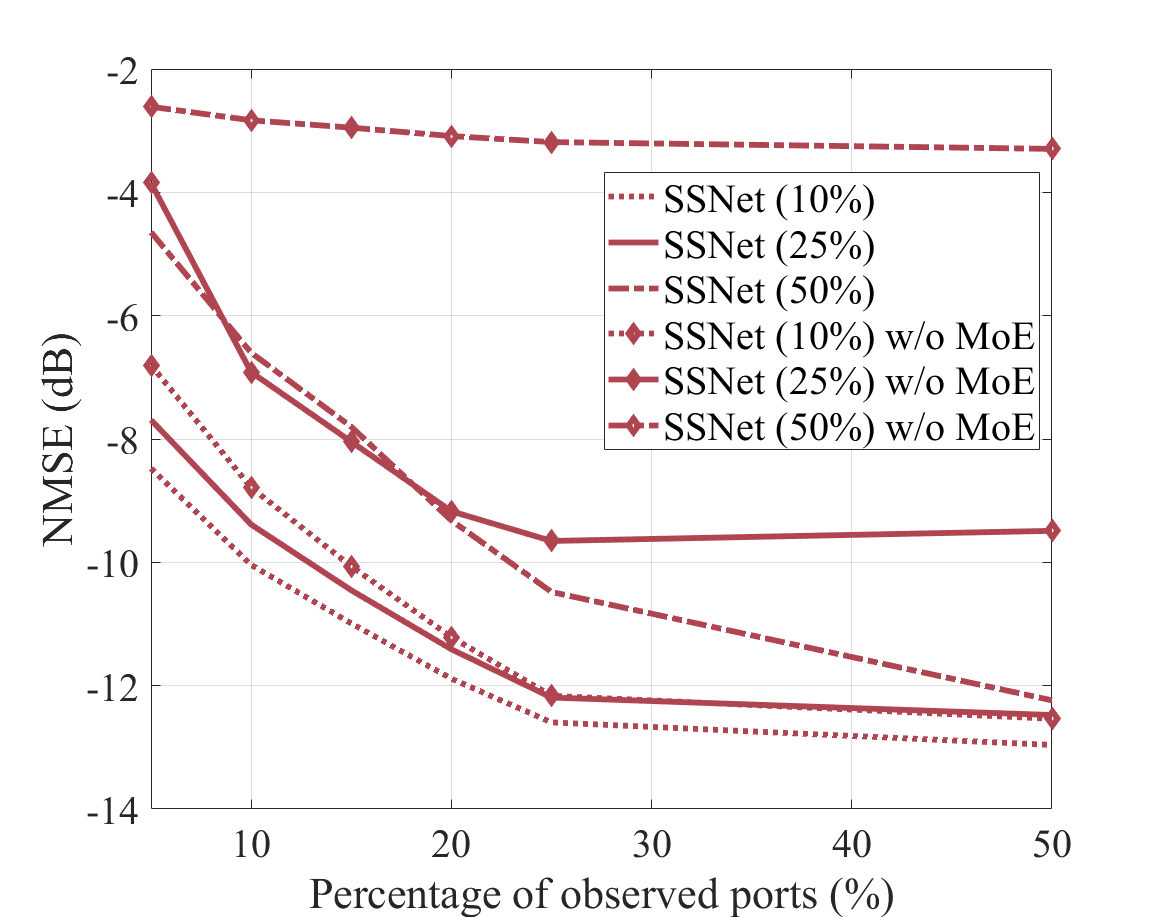} 
}    
\subfigure[20 dB.] { 
\label{ablation_comparison_2_[20db]_[5,10,15,20,25,50]_2_4}    
\includegraphics[width=1.\columnwidth]{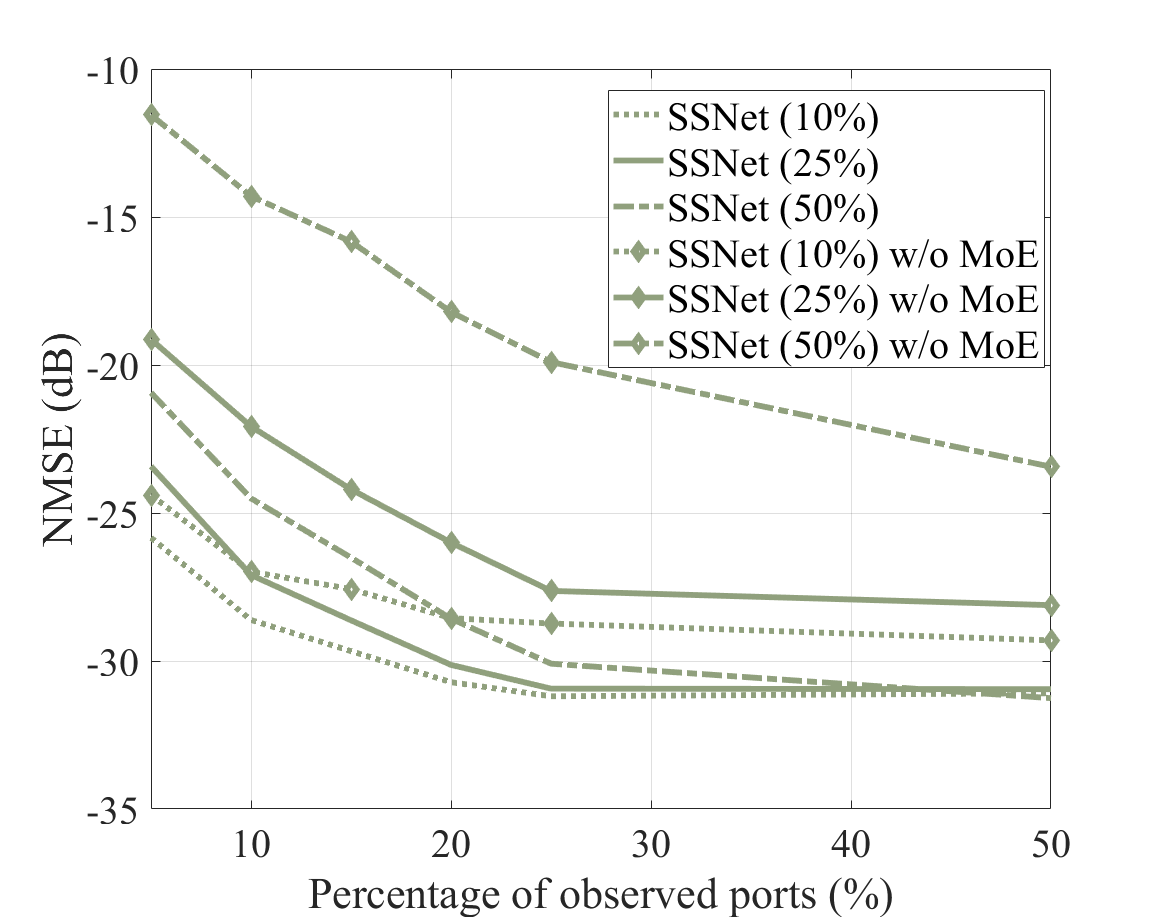} 
}  
\caption{The effect of mask ratio for the proposed SSNet and the ablation experiments to demonstrate the effectiveness of MoE in channel extrapolation. Compared with the proposed SSNet, each encoder block of SSNet w/o MoE, is a stack of an MSA module and an FFN interleaved with residual connections
and LayerNorm.}   
\label{fig:Abliation_experiments}   
\end{figure}

A concern for the performance superiority of SSNet (10\%) over SSNet (25\%) may attribute to that the percentage of observed ports for training and testing of SSNet (10\%) are the same in Fig. \ref{CSI_data_imag_2_4_0dB} and \ref{CSI_data_imag_2_4_20dB}. We evaluate the effect of mask ratio in modeling training comprehensively, we train the proposed SSNet with a mask ratio of 50\%, i.e., SSNet (50\%). We tested SSNet (10\%), SSNet (25\%) and SSNet (50\%) in scenarios with various percentages of observed ports up to 50\%, which is illustrated in Fig. \ref{fig:Abliation_experiments}. 

SSNet (10\%) outperforms SSNet (25\%) and SSNet (50\%) with various percentages of observed ports, including 5\%, 10\%, 15\%, 20\%, 25\% and 50\%. Taking the 0 dB scenario in Fig. \ref{ablation_comparison_2_[0db]_[5,10,15,20,25,50]_2_4}, SSNet (10\%) outperforms SSNet (25\%) and SSNet (50\%) with 5\% of observed ports by 1.2 and 3.8 dB, respectively. By further increasing the percentage of observed ports to 50\%, the performance superiority of SSNet (10\%) over SSNet (25\%) and SSNet (50\%) decrease to 0.4 and 0.9 dB, showing that the channel extrapolation accuracy of SSNet (10\%), SSNet (25\%) and SSNet (50\%) are emerging towards to the same level. Such trend is more obvious in 20 dB sceanrio in Fig. \ref{ablation_comparison_2_[20db]_[5,10,15,20,25,50]_2_4}. 

The above comparison of the proposed models trained with different mask ratios demonstrates that by using a smaller percentage of observed ports during training, the capability of the SSNet is enhanced in testing. This is attributed that by training the model with CSI of a small number of ports, the model is forced to learn the correlation between ports that is more effective for channel extrapolation, thereby showing superior NMSE. As illustrated in Fig. \ref{comparison_loss_[50,75,90]_2_4}, although the training loss of SSNet (10\%), SSNet (25\%) and SSNet (50\%) converage to the same level, approaching 0, they show different patterns during training. After showing a sharp decrease, the training loss SSNet (10\%), SSNet (25\%) and SSNet (50\%) keep nearly unchanged at the level of approximately 0.3 for 31, 18 and 13 epochs, respectively. This indicates that SSNet (10\%) is more difficult to train than SSNet (25\%) and SSNet (50\%), which echos with the above analysis that the superiority of SSNet (10\%) in channel extrapolation is achieved at the expense of training difficulties illustrated.

\subsubsection{Abliation experiments}
To demonstrate the effectiveness of the proposed MoE in channel extrapolation, we carried out ablation experiments. We compare the compare the performance between SSNet and SSNet w/o MoE comprehensively. The MoE module in the endocer of SSNet is replaced by an FFN, i.e., each encoder block of SSNet w/o MoE, is a stack of an MSA module, an FFN interleaved with residual connections and LayerNorm. Fig. \ref{fig:Abliation_experiments} illustrates the NMSE of channel extrapolation between SSNet and SSNet w/o MoE in various cases. A remarkable performance gain of SSNet over SSNet w/o MoE is observed. For example in Fig. \ref{ablation_comparison_2_[20db]_[5,10,15,20,25,50]_2_4}, SSNet (10\%), SSNet (25\%) and SSNet (50\%) outperform SSNet w/o MoE (10\%), SSNet w/o MoE (25\%) and SSNet w/o MoE(50\%) by 4.9 dB, 5.7 dB and 9.1 dB at 20 db with 5\% of port observable. The performance gain reduce to 2.5, 3.6 and 6.8 dB, respectively at 20 db with 50\% of port observable. This indicates that MoE is more effective to learn the channel correlation between ports with a smaller percentage of observed ports. 

\subsubsection{Generalization evaluation}

To rigorously evaluate the generalization capability of SSNet under realistic propagation conditions, we perform zero-shot learning of SSNet on the fully correlated channel model\cite{new2024tutorial}. This advanced model captures directive scattering effects prevalent in urban environments, providing a more challenging testbed compared to the isotropic Clarke's model used during training. The fully correlated model shares a similar structure as Clarke's model, including the eigenvalue decomposition framework in Eq. (\ref{eq:SVG}) and the channel generation framework in Eq. (\ref{eq:channel_generation}).

The core distinction between the Clarke's model and fully correlated model is the spatial correlation computation. The Clarke's model assumes isotropic scattering with uniform AoA distribution in Eq. (\ref{eq:clarke_corr}), while the fully correlated model employs the zero-order Bessel function of the first kind\cite{baricz2010generalized} to model the channel spatial correlation:
\begin{equation}
\Sigma_{\text{full}}(i,j) = J_0\left(2\pi \cdot \|\mathbf{p}_i - \mathbf{p}_j\| / \lambda\right)
\label{eq:bessel_corr}
\end{equation}

This Bessel-function formulation $J_0(\cdot)$ fundamentally alters the spatial correlation characteristics. The Bessel function $J_0(\cdot)$ introduces two critical physical phenomena not captured by Clarke's model: spatial correlation decays slower with antenna separation distance, maintaining significant correlation at larger spacings due to the oscillatory nature of $J_0(z) \sim \sqrt{2/(\pi z)}\cos(z-\pi/4)$ for large $z$\cite{bowman2012introduction}; and directional correlation patterns emerge where spatial correlation varies with orientation relative to dominant scattering paths, as the Bessel solution satisfies the Helmholtz wave equation $\nabla^2 \Sigma + k^2 \Sigma = 0$ with $k=2\pi/\lambda$\cite{olver2009bessel}, inherently encoding directional wave propagation effects absent in Clarke's isotropic model.

A total number of 4,000 channel samples are generated using the above fully correlated channel model. Additive Gaussian noise is added to these samples under predefined SNR for testing, which are illustrated in Fig. \ref{generalization_comparison_2_[0,10,20]_[5,10,15,20,25]_2_4}.The in-distribution case indicates that the proposed SSNet is trained and tested using the channel samples generated using the CLarke's model, while the out-of-distribution case indicates that the proposed SSNet is trained and tested using the channel samples generated using the CLarke's model\cite{bjornson2020rayleigh,ramirez2024new} and fully correlated channel model\cite{new2024tutorial}, respectively.   

We carried out zero-shot learning of the proposed SSNet in the dataset generated using the fully correlated channel model, i.e., the pre-trained SSNet is directly tested using the dataset without changing the weights of the proposed model, such as fine turning. The simulation results are illustrated in Fig. \ref{generalization_comparison_2_[0,10,20]_[5,10,15,20,25]_2_4}, where the in-distribution case indicates that the proposed SSNet is trained and tested using the channel samples generated using the CLarke's model, while the out-of-distribution case indicates that the proposed SSNet is trained and tested using the channel samples generated using the CLarke's model\cite{bjornson2020rayleigh,ramirez2024new} and fully correlated channel model\cite{new2024tutorial}, respectively. Iteratively, there are performance gaps between the in-distribution cases and their out-of-distribution counterparts. For various level of SNRs and percentages of observed ports, the performance loss in out-of-distribution cases are within the range between 3 dB and 5 dB. Such moderate performance loss is manageable, which validates the generalization capability of the proposed model.  

\begin{figure}[htbp]\centering    
\includegraphics[width=1.\columnwidth]{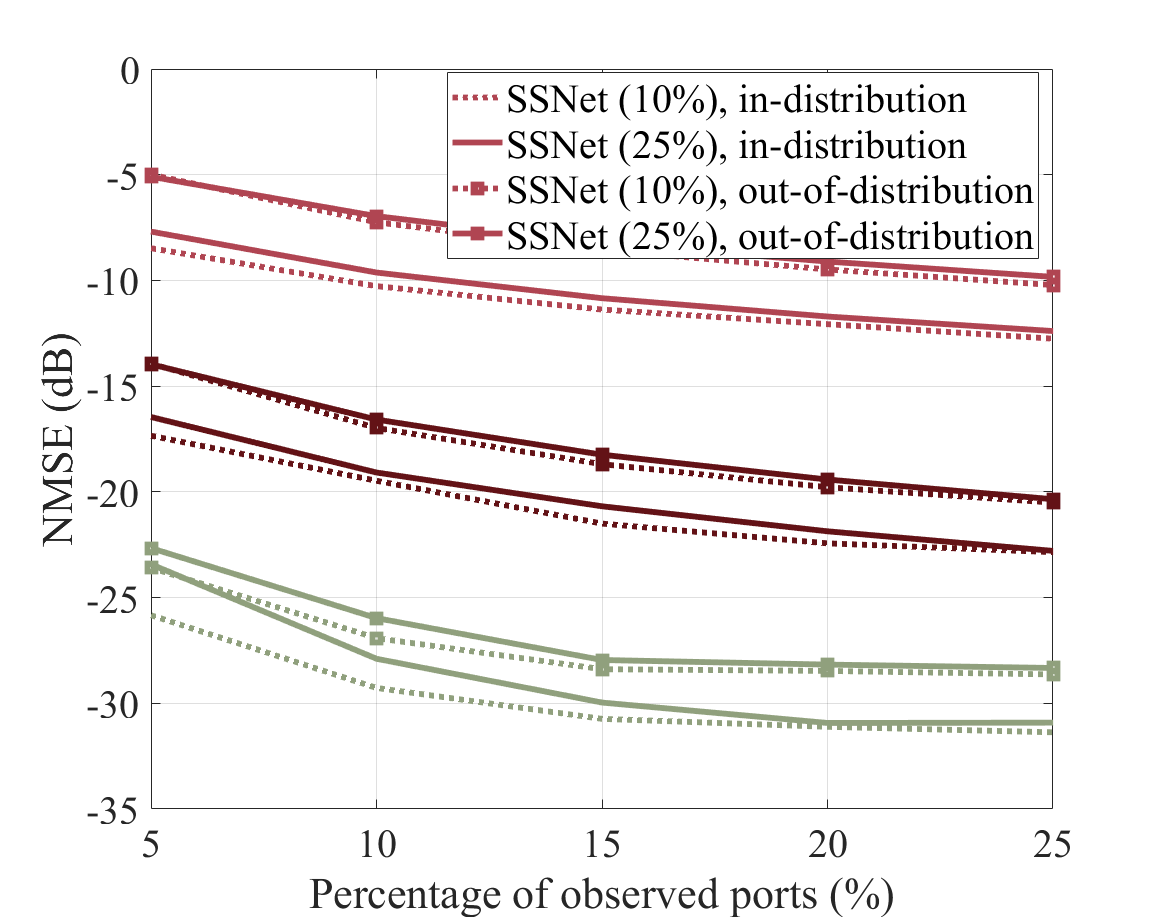} 
\caption{An illustration of the generalization performance. The in-distribution case indicates that the proposed SSNet is trained and tested using the channel samples generated using the CLarke's model, while the out-of-distribution case indicates that the proposed SSNet is trained and tested using the channel samples generated using the CLarke's model\cite{bjornson2020rayleigh,ramirez2024new} and fully correlated channel model\cite{new2024tutorial}, respectively.} 
\label{generalization_comparison_2_[0,10,20]_[5,10,15,20,25]_2_4}\end{figure} 
\subsubsection{Inference speed}
To evaluate the execution speed of the various models, we carried out the inference speed tests in various percentages of observed ratio using three types of GPU, i.e., NVIDIA RTX 4090, 4060 and 3060. The simulation results are demonstrated in Table. \ref{Table:inference_time}. The inference speed of the proposed SSNet, SSNet w/o MoE, AGMAE and LSTM increase are the lowest using NVIDIA RTX 4090, and are the highest using NVIDIA RTX 3060, which is attributed to that computational capability of NVIDIA RTX 4090 is the highest among the three GPUs. It is observed that a higher percentage (25 \%) of observed ports requires longer execution time (than 5 \% of observed ports), which is reasonable because a larger volume of data are being processed. Among all the models, AGMAE is most fast to execute, then comes the LSTM, SSNet w/o MoE and SSNet. SSNet outperforms the AGMAE significantly at the expense of approximately 1.13 ms, 2.9 ms and 3.12 ms longer execution time in NVIDIA RTX 4090, 4060 and 3060, respectively. SSNet w/o outperforms the AGMAE moderately at the expense of approximately 0.45 ms, 2.30 ms and 2.22 ms longer execution time in NVIDIA RTX 4090, 4060 and 3060, respectively, which strikes a good balance between the performance and inference time. 
\begin{table*}\label{Table:inference_speed}\centering
\caption{The comparison of various models in terms of inference time (ms) per sample in different hardware platforms and different percentages of observed ports.}
\begin{tabular}{|c|cc|cc|cc|}
\hline
\multirow{2}{*}{} & \multicolumn{2}{c|}{NVIDIA GeForce RTX 4090} & \multicolumn{2}{c|}{NVIDIA GeForce RTX 4060} & \multicolumn{2}{c|}{NVIDIA GeForce RTX 3060} \\ \cline{2-7} 
                     & \multicolumn{1}{c|}{5\%}  & 25\% & \multicolumn{1}{c|}{5\%}   & 25\%  & \multicolumn{1}{c|}{5\%}   & 25\%  \\ \hline
LSTM                 & \multicolumn{1}{c|}{3.02} & 3.01 & \multicolumn{1}{c|}{10.09} & 10.67 & \multicolumn{1}{c|}{12.11} & 12.35 \\ \hline
AGMAE                & \multicolumn{1}{c|}{2.75} & 2.82 & \multicolumn{1}{c|}{9.52}  & 9.67  & \multicolumn{1}{c|}{11.32} & 11.45 \\ \hline
SSNet w/o MoE (90\%) & \multicolumn{1}{c|}{3.23} & 3.29 & \multicolumn{1}{c|}{11.80} & 11.91 & \multicolumn{1}{c|}{13.55} & 13.96 \\ \hline
SSNet w/o MoE (75\%) & \multicolumn{1}{c|}{3.31} & 3.35 & \multicolumn{1}{c|}{11.41} & 11.72 & \multicolumn{1}{c|}{12.96} & 13.12 \\ \hline
SSNet (90\%)         & \multicolumn{1}{c|}{3.90} & 3.98 & \multicolumn{1}{c|}{12.07} & 12.44 & \multicolumn{1}{c|}{14.55} & 14.66 \\ \hline
SSNet (75\%)         & \multicolumn{1}{c|}{3.88} & 4.06 & \multicolumn{1}{c|}{12.40} & 12.53 & \multicolumn{1}{c|}{14.49} & 14.78 \\ \hline
\end{tabular}\label{Table:inference_time}
\end{table*}
\section{Conclusions}
\label{sec:con}
This paper propose SSNet, a novel self-supervised learning framework for channel extrapolation in FAS. To further improve the feature extraction capability and the noise robustness of the self-supervised learning, we propose to train the SSNet using a joint loss function of contrastive loss and channel extrapolation normalized mean squared error. Our findings demonstrate that SSNet significantly outperforms existing methods in terms of accuracy, robustness to noise, and flexibility in handling variable numbers of CSI inputs. Specifically, SSNet outperforms benchmark model with only 5\% data, demonstrating its efficiency in data utilization. The improved robustness, evident in its performance across diverse SNR conditions and FAS sizes, highlights the benefits of the self-supervised learning approach and the joint loss function in mitigating the challenges posed by noise and high-dimensional data. 

However, several key areas require further investigation. While SSNet demonstrates impressive performance in offline settings, the development of a real-time channel extrapolation system is crucial for practical FAS deployment. Future research should focus on optimizing SSNet’s architecture and inference speed to enable real-time CSI estimation. Moreover, the optimal training strategy of SSNet requires further research, for example, what is the optimal percentage of observed ports for training, does a mix of percentage of observed ports for training improve the performance of SSNet. Additional, exploring the integration of SSNet with advanced signal processing techniques, such as beamforming and precoding, will be essential to fully realize the potential of FAS. Finally, extending SSNet to handle more complex scenarios, such as those involving ISAC, presents an exciting direction for future research. The combination of SSNet's efficient channel extrapolation with advanced signal processing techniques holds great promise for unlocking the full potential of FAS in future 6G networks.

\ifCLASSOPTIONcaptionsoff
  \newpage
\fi

\bibliographystyle{IEEEtran}
\bibliography{bibfile}

\begin{thebibliography}{10}
\providecommand{\url}[1]{#1}
\csname url@samestyle\endcsname
\providecommand{\newblock}{\relax}
\providecommand{\bibinfo}[2]{#2}
\providecommand{\BIBentrySTDinterwordspacing}{\spaceskip=0pt\relax}
\providecommand{\BIBentryALTinterwordstretchfactor}{4}
\providecommand{\BIBentryALTinterwordspacing}{\spaceskip=\fontdimen2\font plus
\BIBentryALTinterwordstretchfactor\fontdimen3\font minus \fontdimen4\font\relax}
\providecommand{\BIBforeignlanguage}[2]{{%
\expandafter\ifx\csname l@#1\endcsname\relax
\typeout{** WARNING: IEEEtran.bst: No hyphenation pattern has been}%
\typeout{** loaded for the language `#1'. Using the pattern for}%
\typeout{** the default language instead.}%
\else
\language=\csname l@#1\endcsname
\fi
#2}}
\providecommand{\BIBdecl}{\relax}
\BIBdecl

\bibitem{wong2022bruce}
K.-K. Wong, K.-F. Tong, Y.~Shen, Y.~Chen, and Y.~Zhang, ``Bruce lee-inspired fluid antenna system: {Six} research topics and the potentials for {6G},'' \emph{Frontiers in Communications and Networks}, vol.~3, p. 853416, 2022.

\bibitem{new2024tutorial}
W.~K. New, K.-K. Wong, H.~Xu, C.~Wang, F.~R. Ghadi, J.~Zhang, J.~Rao, R.~Murch, P.~Ram{\'\i}rez-Espinosa, D.~Morales-Jimenez \emph{et~al.}, ``A tutorial on fluid antenna system for 6g networks: {Encompassing} communication theory, optimization methods and hardware designs,'' \emph{IEEE Communications Surveys \& Tutorials}, 2024.

\bibitem{faddoul2024advanced}
E.~Faddoul, G.~M. Kraidy, C.~Psomas, and I.~Krikidis, ``Advanced channel coding designs for index-modulated fluid antenna systems,'' \emph{arXiv preprint arXiv:2403.06839}, 2024.

\bibitem{zou2024shifting}
J.~Zou, H.~Xu, C.~Wang, L.~Xu, S.~Sun, K.~Meng, C.~Masouros, and K.-K. Wong, ``Shifting the {ISAC} trade-off with fluid antenna systems,'' \emph{arXiv preprint arXiv:2405.05715}, 2024.

\bibitem{wong2024virtual}
K.-K. Wong, C.~Wang, H.~Zhang, G.~Li, C.-C. Wang, C.-B. Chae, and R.~Murch, ``Virtual {FAS} by learning-based imaginary antennas,'' \emph{IEEE Wireless Communications Letters}, 2024.

\bibitem{xu2024revisiting}
H.~Xu, K.-K. Wong, W.~K. New, K.-F. Tong, Y.~Zhang, and C.-B. Chae, ``Revisiting outage probability analysis for two-user fluid antenna multiple access system,'' \emph{IEEE Transactions on Wireless Communications}, 2024.

\bibitem{new2023fluid}
W.~K. New, K.-K. Wong, H.~Xu, K.-F. Tong, and C.-B. Chae, ``Fluid antenna system: {New} insights on outage probability and diversity gain,'' \emph{IEEE Transactions on Wireless Communications}, vol.~23, no.~1, pp. 128--140, 2023.

\bibitem{psomas2023diversity}
C.~Psomas, G.~M. Kraidy, K.-K. Wong, and I.~Krikidis, ``On the diversity and coded modulation design of fluid antenna systems,'' \emph{IEEE Transactions on Wireless Communications}, 2023.

\bibitem{xu2023channel}
H.~Xu, G.~Zhou, K.-K. Wong, W.~K. New, C.~Wang, C.-B. Chae, R.~Murch, S.~Jin, and Y.~Zhang, ``Channel estimation for {FAS}-assisted multiuser {mmWave} systems,'' \emph{IEEE Communications Letters}, 2023.

\bibitem{new2024channel}
W.~K. New, K.-K. Wong, H.~Xu, F.~R. Ghadi, R.~Murch, and C.-B. Chae, ``Channel estimation and reconstruction in fluid antenna system: {Oversampling} is essential,'' \emph{arXiv preprint arXiv:2405.15607}, 2024.

\bibitem{skouroumounis2022fluid}
C.~Skouroumounis and I.~Krikidis, ``Fluid antenna with linear {MMSE} channel estimation for large-scale cellular networks,'' \emph{IEEE Transactions on Communications}, vol.~71, no.~2, pp. 1112--1125, 2022.

\bibitem{zhang2023successive}
Z.~Zhang, J.~Zhu, L.~Dai, and R.~W. Heath, ``Successive {Bayesian} reconstructor for channel estimation in fluid antenna systems,'' \emph{IEEE Transactions on Wireless Communications}, 2024.

\bibitem{jin2025linformer}
Y.~Jin, Y.~Wu, Y.~Gao, S.~Zhang, S.~Xu, and C.-X. Wang, ``Linformer: {A} linear-based lightweight transformer architecture for time-aware {MIMO} channel prediction,'' \emph{IEEE Transactions on Wireless Communications}, 2025.

\bibitem{zhang2021deep}
S.~Zhang, Y.~Liu, F.~Gao, C.~Xing, J.~An, and O.~A. Dobre, ``Deep learning based channel extrapolation for large-scale antenna systems: {Opportunities}, challenges and solutions,'' \emph{IEEE Wireless Communications}, vol.~28, no.~6, pp. 160--167, 2021.

\bibitem{zhang2023ai}
Z.~Zhang, J.~Zhang, Y.~Zhang, L.~Yu, and G.~Liu, ``{AI}-based time-, frequency-, and space-domain channel extrapolation for {6G}: {Opportunities} and challenges,'' \emph{IEEE Vehicular Technology Magazine}, vol.~18, no.~1, pp. 29--39, 2023.

\bibitem{qiu2024can}
Y.~Qiu, D.~Wu, and Y.~Zeng, ``Can channels be fully inferred between two antenna panels?'' \emph{IEEE Communications Letters}, vol.~28, no.~4, pp. 942--946, 2024.

\bibitem{liu2022massive}
W.~Liu, Z.~Chen, and X.~Gao, ``Massive {MIMO} channel prediction in real propagation environments using tensor decomposition and autoregressive models,'' in \emph{2022 IEEE 33rd Annual International Symposium on Personal, Indoor and Mobile Radio Communications (PIMRC)}.\hskip 1em plus 0.5em minus 0.4em\relax IEEE, 2022, pp. 849--855.

\bibitem{rottenberg2019channel}
F.~Rottenberg, R.~Wang, J.~Zhang, and A.~F. Molisch, ``Channel extrapolation in {FDD} massive {MIMO}: {Theoretical} analysis and numerical validation,'' in \emph{2019 IEEE Global Communications Conference (GLOBECOM)}.\hskip 1em plus 0.5em minus 0.4em\relax IEEE, 2019, pp. 1--7.

\bibitem{han2019efficient}
Y.~Han, T.-H. Hsu, C.-K. Wen, K.-K. Wong, and S.~Jin, ``Efficient downlink channel reconstruction for {FDD} multi-antenna systems,'' \emph{IEEE Transactions on Wireless Communications}, vol.~18, no.~6, pp. 3161--3176, 2019.

\bibitem{jin2023model}
W.~Jin, J.~Zhang, C.-K. Wen, and S.~Jin, ``Model-driven deep learning for hybrid precoding in millimeter wave {MU-MIMO} system,'' \emph{IEEE Transactions on Communications}, 2023.

\bibitem{zhang2022deep}
S.~Zhang, S.~Zhang, J.~Ma, T.~Liu, and O.~A. Dobre, ``Deep learning based antenna-time domain channel extrapolation for hybrid mmwave massive {MIMO},'' \emph{IEEE Transactions on Vehicular Technology}, vol.~71, no.~12, pp. 13\,398--13\,402, 2022.

\bibitem{xu2022sparse}
X.~Xu, S.~Zhang, F.~Gao, and J.~Wang, ``Sparse bayesian learning based channel extrapolation for {RIS} assisted {MIMO-OFDM},'' \emph{IEEE Transactions on Communications}, vol.~70, no.~8, pp. 5498--5513, 2022.

\bibitem{gao2025joint}
Y.~Gao, X.~Xu, Y.~Jin, W.~Yuan, J.~Zhang, and S.~Xu, ``Joint channel estimation and data detection for {OTFS} systems: {A} lightweight deep learning framework with a novel data augmentation method,'' \emph{IEEE Internet of Things Journal}, 2025.

\bibitem{yu2024multi}
W.~Yu, J.~Jiang, Y.~Gao, and S.~Xu, ``Multi-scenario time-domain channel extraploation: {A} {Transformer}-based approach,'' in \emph{2024 IEEE 24th International Conference on Communication Technology (ICCT)}.\hskip 1em plus 0.5em minus 0.4em\relax IEEE, 2024, pp. 1446--1450.

\bibitem{jiang2025towards}
J.~Jiang, Y.~Gao, X.~Wu, and S.~Xu, ``Towards channel foundation models ({CFMs}): {Motivations}, methodologies and opportunities,'' \emph{arXiv preprint arXiv:2507.13637}, 2025.

\bibitem{gao2025enabling}
Y.~Gao, Z.~Lu, Y.~Wu, Y.~Jin, S.~Zhang, X.~Chu, S.~Xu, and C.-X. Wang, ``Enabling {6G} through multi-domain channel extrapolation: {Opportunities} and challenges of generative artificial intelligence,'' \emph{arXiv preprint arXiv:2509.01125}, 2025.

\bibitem{lin2021deep}
B.~Lin, F.~Gao, S.~Zhang, T.~Zhou, and A.~Alkhateeb, ``Deep learning-based antenna selection and {CSI} extrapolation in massive {MIMO} systems,'' \emph{IEEE Transactions on Wireless Communications}, vol.~20, no.~11, pp. 7669--7681, 2021.

\bibitem{waqar2023deep}
N.~Waqar, K.-K. Wong, K.-F. Tong, A.~Sharples, and Y.~Zhang, ``Deep learning enabled slow fluid antenna multiple access,'' \emph{IEEE Communications Letters}, vol.~27, no.~3, pp. 861--865, 2023.

\bibitem{zhang2024learning}
H.~Zhang, J.~Wang, C.~Wang, C.-C. Wang, K.-K. Wong, B.~Wang, and C.-B. Chae, ``Learning-induced channel extrapolation for fluid antenna systems using asymmetric graph masked autoencoder,'' \emph{IEEE Wireless Communications Letters}, 2024.

\bibitem{wu2024channel}
X.~Wu, H.~Zhang, C.-C. Wang, and Z.~Li, ``Channel state information extrapolation in fluid antenna systems based on masked language model,'' in \emph{2024 IEEE International Conference on Communications Workshops (ICC Workshops)}.\hskip 1em plus 0.5em minus 0.4em\relax IEEE, 2024, pp. 1383--1388.

\bibitem{wang2024fluid}
C.~Wang, G.~Li, H.~Zhang, K.-K. Wong, Z.~Li, D.~W.~K. Ng, and C.-B. Chae, ``Fluid antenna system liberating multiuser {MIMO} for {ISAC }via deep reinforcement learning,'' \emph{IEEE Transactions on Wireless Communications}, 2024.

\bibitem{jin2025dual}
Y.~Jin, P.~Qi, Y.~Gao, and S.~Liu, ``Dual-path residual attention network for efficient channel estimation in {RIS}-assisted communication systems,'' \emph{Physical Communication}, vol.~68, p. 102577, 2025.

\bibitem{gao2024performance}
Y.~Gao, H.~Hu, J.~Zhang, Y.~Jin, S.~Xu, and X.~Chu, ``On the performance of an integrated communication and localization system: {An} analytical framework,'' \emph{IEEE Transactions on Vehicular Technology}, vol.~73, no.~7, pp. 10\,845--10\,849, 2024.

\bibitem{jiang2025c2s}
J.~Jiang, S.~Xu, W.~Yu, and Y.~Gao, ``{C2S-AE}: {CSI} to sensing enabled by an auto-encoder-based framework,'' \emph{arXiv preprint arXiv:2503.00941}, 2025.

\bibitem{gao2024c2s}
Y.~Gao, X.~Wu, Y.~Gao, and S.~Xu, ``{C2S}: {An} {Transformer}-based framework to extrapolate sensing channel from communication channel,'' in \emph{2024 IEEE 100th Vehicular Technology Conference (VTC2024-Fall)}.\hskip 1em plus 0.5em minus 0.4em\relax IEEE, 2024, pp. 1--5.

\bibitem{li2024model}
G.~Li, H.~Zhang, C.~Wang, and B.~Wang, ``Model-driven channel extrapolation for massive fluid antenna,'' in \emph{2024 IEEE International Conference on Communications Workshops (ICC Workshops)}.\hskip 1em plus 0.5em minus 0.4em\relax IEEE, 2024, pp. 1146--1151.

\bibitem{jin2024hsgan}
Y.~Jin, J.~Zhou, and Y.~Gao, ``{HSGAN-IoT}: {A} hierarchical semi-supervised generative adversarial networks for iot device classification,'' \emph{Computer Networks}, vol. 243, p. 110299, 2024.

\bibitem{jin2023sscmt}
Y.~Jin, J.~Fang, and Y.~Gao, ``{SSCMT-ETC}: {A} semi-supervised contrastive mean teacher model for encrypted traffic classification,'' in \emph{2023 2nd International Conference on Sensing, Measurement, Communication and Internet of Things Technologies (SMC-IoT)}.\hskip 1em plus 0.5em minus 0.4em\relax IEEE, 2023, pp. 127--133.

\bibitem{jin2023multiclass}
Y.~Jin, X.~Yu, and Y.~Gao, ``Multiclass malicious {URL} attack type detection via capsule-based neural network,'' in \emph{Third International Seminar on Artificial Intelligence, Networking, and Information Technology (AINIT 2022)}, vol. 12587.\hskip 1em plus 0.5em minus 0.4em\relax SPIE, 2023, pp. 520--525.

\bibitem{van2024generative}
N.~Van~Huynh, J.~Wang, H.~Du, D.~T. Hoang, D.~Niyato, D.~N. Nguyen, D.~I. Kim, and K.~B. Letaief, ``Generative {AI} for physical layer communications: {A} survey,'' \emph{IEEE Transactions on Cognitive Communications and Networking}, vol.~10, no.~3, pp. 706--728, 2024.

\bibitem{xu2025enhanced}
S.~Xu, J.~Jiang, W.~Yu, Y.~Gao, G.~Pan, S.~Mu, Z.~Ai, Y.~Gao, P.~Jiang, and C.-X. Wang, ``Enhanced fingerprint-based positioning with practical imperfections: {Deep} learning-based approaches,'' \emph{arXiv preprint arXiv:2509.01197}, 2025.

\bibitem{zhang2023self}
Z.~Zhang, T.~Ji, H.~Shi, C.~Li, Y.~Huang, and L.~Yang, ``A self-supervised learning-based channel estimation for {IRS}-aided communication without ground truth,'' \emph{IEEE Transactions on Wireless Communications}, vol.~22, no.~8, pp. 5446--5460, 2023.

\bibitem{gui2024survey}
J.~Gui, T.~Chen, J.~Zhang, Q.~Cao, Z.~Sun, H.~Luo, and D.~Tao, ``A survey on self-supervised learning: {Algorithms}, applications, and future trends,'' \emph{IEEE Transactions on Pattern Analysis and Machine Intelligence}, 2024.

\bibitem{wang2024csi}
Y.~Wang, G.~Yu, Y.~Zhang, D.~Liu, and Y.~Zhang, ``{CSI}-based location-independent human activity recognition by contrast between dual stream fusion features,'' \emph{IEEE Sensors Journal}, 2024.

\bibitem{ferrand2023wireless}
P.~Ferrand, M.~Guillaud, C.~Studer, and O.~Tirkkonen, ``Wireless channel charting: {Theory}, practice, and applications,'' \emph{IEEE Communications Magazine}, vol.~61, no.~6, pp. 124--130, 2023.

\bibitem{stephan2024channel}
P.~Stephan, F.~Euchner, and S.~t. Brink, ``Channel charting-based channel prediction on real-world distributed massive {MIMO} {CSI},'' \emph{arXiv preprint arXiv:2410.11486}, 2024.

\bibitem{dai2023vision}
M.~Dai, E.~Zheng, Z.~Feng, L.~Qi, J.~Zhuang, and W.~Yang, ``Vision-based {UAV} self-positioning in low-altitude urban environments,'' \emph{IEEE Transactions on Image Processing}, 2023.

\bibitem{oquab2023dinov2}
M.~Oquab, T.~Darcet, T.~Moutakanni, H.~Vo, M.~Szafraniec, V.~Khalidov, P.~Fernandez, D.~Haziza, F.~Massa, A.~El-Nouby \emph{et~al.}, ``Dinov2: {Learning} robust visual features without supervision,'' \emph{arXiv preprint arXiv:2304.07193}, 2023.

\bibitem{he2022masked}
K.~He, X.~Chen, S.~Xie, Y.~Li, P.~Doll{\'a}r, and R.~Girshick, ``Masked autoencoders are scalable vision learners,'' in \emph{Proceedings of the IEEE/CVF conference on computer vision and pattern recognition}, 2022, pp. 16\,000--16\,009.

\bibitem{vaswani2017attention}
A.~Vaswani, N.~Shazeer, N.~Parmar, J.~Uszkoreit, L.~Jones, A.~N. Gomez, {\L}.~Kaiser, and I.~Polosukhin, ``Attention is all you need,'' \emph{Advances in neural information processing systems}, vol.~30, 2017.

\bibitem{rani2023self}
V.~Rani, S.~T. Nabi, M.~Kumar, A.~Mittal, and K.~Kumar, ``Self-supervised learning: {A} succinct review,'' \emph{Archives of Computational Methods in Engineering}, vol.~30, no.~4, pp. 2761--2775, 2023.

\bibitem{bjornson2020rayleigh}
E.~Bj{\"o}rnson and L.~Sanguinetti, ``Rayleigh fading modeling and channel hardening for reconfigurable intelligent surfaces,'' \emph{IEEE Wireless Communications Letters}, vol.~10, no.~4, pp. 830--834, 2020.

\bibitem{ramirez2024new}
P.~Ram{\'\i}rez-Espinosa, D.~Morales-Jimenez, and K.-K. Wong, ``A new spatial block-correlation model for fluid antenna systems,'' \emph{IEEE Transactions on Wireless Communications}, 2024.

\bibitem{lee2023mathematical}
M.~Lee, ``Mathematical analysis and performance evaluation of the {GeLu} activation function in deep learning,'' \emph{Journal of Mathematics}, vol. 2023, no.~1, p. 4229924, 2023.

\bibitem{pinkus1999approximation}
A.~Pinkus, ``Approximation theory of the {MLP} model in neural networks,'' \emph{Acta numerica}, vol.~8, pp. 143--195, 1999.

\bibitem{baricz2010generalized}
{\'A}.~Baricz, \emph{Generalized Bessel functions of the first kind}.\hskip 1em plus 0.5em minus 0.4em\relax Springer, 2010.

\bibitem{bowman2012introduction}
F.~Bowman, \emph{Introduction to Bessel functions}.\hskip 1em plus 0.5em minus 0.4em\relax Courier Corporation, 2012.

\bibitem{olver2009bessel}
F.~W.~J. Olver, L.~C. Maximon, D.~Lozier, R.~Boisvert, and C.~Clark, ``Bessel functions,'' \emph{NIST handbook of mathematical functions}, no. 2655350, pp. 215--286, 2009.

\end{thebibliography}

\end{document}